\documentclass[journal]{IEEEtran}
\usepackage{graphicx}
\usepackage{cite}
\usepackage{picinpar}
\usepackage{amsmath}
\usepackage{url}
\usepackage{flushend}
\usepackage[latin1]{inputenc}
\usepackage{colortbl}
\usepackage{soul}
\usepackage{multirow}
\usepackage{pifont}
\usepackage{color}
\usepackage{alltt}
\usepackage[hidelinks]{hyperref}
\usepackage{enumerate}
\usepackage{siunitx}
\usepackage{breakurl}
\usepackage{epstopdf}
\usepackage{pbox}
\usepackage{makecell}
\usepackage{amssymb}
\usepackage{algorithm}
\usepackage{algorithmic}
\usepackage{mathrsfs}
\ifCLASSINFOpdf
\else
\fi
\begin{document}
%
\title{Decision Making for Connected Automated Vehicles at Urban Intersections Considering Social and Individual Benefits}
%
%

\author{Peng Hang,
        Chao Huang, Zhongxu Hu, and Chen Lv
\thanks{This work was supported in part by  the  A*STAR  Grant (No.1922500046), and the SUG-NAP Grant (No.M4082268.050) of Nanyang Technological University, Singapore.}
\thanks{P. Hang, C. Huang, Z. Hu and C. Lv, are with the School of Mechanical and Aerospace Engineering, Nanyang Technological University, Singapore 639798. (e-mail: \{peng.hang, chao.huang, zhongxu.hu, lyuchen\}@ntu.edu.sg)}
\thanks{Corresponding author: C. Lv}}

%
%

\markboth{ }
{Shell \MakeLowercase{\textit{et al.}}: }
%



\maketitle

\begin{abstract}
To address the coordination issue of connected automated vehicles (CAVs) at urban scenarios, a game-theoretic decision-making framework is proposed that can advance social benefits, including the traffic system efficiency and safety, as well as the benefits of individual users. Under the proposed decision-making framework, in this work, a representative urban driving scenario, i.e. the unsignalized intersection, is investigated. Once the vehicle enters the focused zone, it will interact with other CAVs and make collaborative decisions. To evaluate the safety risk of surrounding vehicles and reduce the complexity of the decision-making algorithm, the driving risk assessment algorithm is designed with a Gaussian potential field approach. The decision-making cost function is constructed by considering the driving safety and passing efficiency of CAVs. Additionally, decision-making constraints are designed and include safety, comfort, efficiency, control and stability. Based on the cost function and constraints, the fuzzy coalitional game approach is applied to the decision-making issue of CAVs at unsignalized intersections. Two types of fuzzy coalitions are constructed that reflect both individual and social benefits. The benefit allocation in the two types of fuzzy coalitions is associated with the driving aggressiveness of CAVs. Finally, the effectiveness and feasibility of the proposed decision-making framework are verified with three test cases.

\end{abstract}


\IEEEpeerreviewmaketitle

\section{Introduction}
\subsection{Motivation}
\IEEEPARstart{I}{n} existing traffic systems, traffic signal control is an effective method to address traffic congestion or conflict and advance the driving safety of vehicles, especially at crossroad intersections \cite{zhao2018platoon}. However, the application of traffic signal control also leads to a reduction in traffic travel efficiency. With the development of intelligent traffic systems and automated driving techniques, traffic signal control is not necessary, especially for connected automated vehicles (CAVs) \cite{bichiou2018developing,zhang2020trajectory}. The entering vehicle does not require a complete stop to wait for the green light. As a result, traffic delays are reduced, and traffic capacity is improved \cite{tajalli2018distributed,hang2020cooperative}. Therefore, connected and automated driving techniques are a favorable vantage point for addressing the issue of traffic congestion and conflict at unsignalized intersections to improve traffic efficiency and advance vehicle driving safety \cite{okumura2016challenges}. In the connected driving environment, all kinds of information, including the vehicle motion state, driving environment information, and even driving intentions and behaviors, can be shared with each entity, which will enhance the intelligent decision-making of CAVs. As a result, the driving performances of CAVs can be improved to include safety, comfort, efficiency and fuel economy \cite{hang2020human}.

\subsection{Related Work}
In recent years, the centralized traffic management approach has been widely studied to address the driving conflict of CAVs at unsignalized intersections. All CAVs that enter the controlled zone must hand over their control authorities to the centralized controller. Namely, each CAV has no authority to conduct the individual their self-driving, and they are controlled by the centralized controller. Then, the centralized traffic management system will manage the scheduling of all vehicles in the control zone according to the optimal passing sequence \cite{dresner2008multiagent,carlino2013auction}. The model-based approach has been widely used for the centralized traffic management. With the application of the agent-based traffic control method, an autonomous intersection management (AIM) system was proposed to improve the traffic travel efficiency \cite{hausknecht2011autonomous}. In \cite{bouderba2019v2x}, based on the right-hand priority, a cooperative intersection management system was designed to enhance the average vehicle velocity while retaining a high capacity of the intersection. To ensure driving safety and improve the travel efficiency for CAVs at unsignalized intersections, an alternately iterative descent method (AIDM) was applied to assign the optimal entry time for each CAV \cite{qian2019toward}. In addition to the model-based approach, the data-driven approach is also applied to the centralized traffic management. In \cite{tallapragada2015coordinated}, an intersection manager was designed with the cluster approach to optimize the passing order and optimal velocities for CAVs at unsignalized intersections, which can minimize the total time and energy consumption. In \cite{wang2020cooperative}, a cooperative autonomous traffic organization method was studied for CAVs in a multi-intersection road network, which showed advantages over conventional baseline schemes in terms of global traffic efficiency.

Although the centralized traffic management system can improve traffic efficiency, the computational complexity increases with the increase of vehicles. In addition, it considers the optimization from the point of the entire traffic system, and the individual benefit of each CAV, e.g., personalized decision making, is ignored \cite{bian2019cooperation}. Distributed conflict resolution is another common method to address the decision-making issue for CAVs at unsignalized intersections, in which CAVs can realize collaborative decision making without the help of a centralized traffic manager. Four kinds of approaches have been applied to the distributed conflict resolution, including the rule-based approach, the model-based approach, the data-driven approach and the game theoretic approach. The rule-based approach is the simplest. With the combination of Monte Carlo tree search (MCTS) and some heuristic rules, a new cooperative decision-making strategy was designed for CAVs at unsignalized intersections, which can maintain a good trade-off between performance and computational flexibility \cite{xu2019cooperative}.
To improve the reliability, robustness, safety and efficiency of CAVs at intersections, a digital map is used in the decision-making framework to predict the paths of surrounding vehicles; then, potential threats are provided to the ego vehicle to make safe and efficient decisions \cite{noh2018decision}. As to the data-driven approach, a Q-learning approach was proposed to address the decision-making issue of CAVs at unsignalized intersections based on single-agent Q-learning and Nash equilibrium \cite{guo2020evaluating}. In \cite{gadginmath2020data}, a data-driven approach is applied to the distributed intersection management for CAVs, in which the intersection usage sequence is obtained through a data-driven online classification and the trajectories are computed sequentially. The performance of the data-driven approach depends on the quantity and quality of the dataset.
Additionally, the model-based approaches are widely used.
In \cite{debada2018virtual}, based on a virtual vehicle-based interaction mechanism and traffic prediction, a distributed maneuver planner was designed for CAVs to address decision making and trajectory planning at a roundabout intersection. In \cite{liu2017distributed}, a novel communication-enabled distributed conflict resolution mechanism was proposed for CAVs, which favored a group of CAVs to navigate safely and efficiently in intersections without any traffic manager. To enhance driving safety and travel efficiency simultaneously at unsignalized intersections, a combined spring model for assessing driving risk was applied to the decision-making algorithm for CAVs \cite{zheng2021behavioral}. In \cite{cho2018real}, the obstacle-dependent Gaussian potential field (ODG-PF) approach is applied to the motion planning and decision making for collision avoidance.

Besides, the game theoretic-based approach shows superiority and effectiveness in modeling the interaction and decision making of CAVs \cite{hang2021decision}. In general, game theoretic-based approaches for decision making can be divided into two types, i.e., the noncooperative game theoretic-based approach and the cooperative game theoretic-based approach. The noncooperative game theoretic-based approach, such as the Stackelberg game, is usually applied to the interaction modelling and decision making between two or more players under a driving environment with incomplete information \cite{hang2020human}. As a result, a compromise equilibrium solution can be accepted by all players. However, it is not the global optimal solution. The cooperative game theoretic-based approach, such as the grand coalitional game, is usually used for the interaction modelling and decision making of multiple players under a driving environment with complete information, e.g., connected driving environment for vehicles \cite{hang2021cooperative}. In the cooperative game issue, players can obtain a global optimal solution with the collaboration with each other. In \cite{rahmati2017towards}, a noncooperative game approach was applied to unprotected left-turn decision making of CAVs at unsignalized intersections, which can effectively capture vehicle interactions and make human-like decisions for CAVs.
The fuzzy coalitional game approach is a typical cooperative game approach which is effective to deal with the conflict between multiple agents \cite{aubin1981cooperative}. A cooperative game with fuzzy core is used to form a coalition allowing coordinating the actions of individual members to achieve a common goal, as well as to evaluate and distribute the overall benefit \cite{smirnov2019fuzzy}.

Based on the above literature review, it can be found that existing studies usually consider the social benefit and the individual benefit separately. The social benefit is the benefit for the entire traffic system, including traffic efficiency and safety, which is the collective benefit for all CAVs. The individual benefit is the interest for each single CAV. Only social benefit is considered in the centralized traffic management system, and only individual benefit is considered in the distributed conflict resolution algorithm. To advance the performance of the entire traffic system and meanwhile care the interest of the individual, both social benefit and individual benefit are considered in the decision-making algorithm design for CAVs in this paper.

\subsection{Contribution}
Comprehensively considering social and individual benefits at unsignalized intersections, a game-theoretic decision-making framework is designed for CAVs. The contributions of this paper are summarized as follows. (1) Based on a Gaussian potential field approach, a driving risk assessment algorithm is proposed to evaluate safety risks of surrounding vehicles and increase the computational efficiency of the decision-making algorithm. (2) In the decision-making model, both driving safety and passing efficiency are considered. The personalized decision-making feature for CAVs is reflected by introducing the driving aggressiveness into the model. Besides, multiple performance constraints are considered in the decision-making algorithm, including safety, comfort, efficiency, control and stability. (3) The fuzzy coalitional game approach is utilized to address the decision-making issue of CAVs at unsignalized intersections, which is in favor of the individual benefits of CAVs and the social benefit of the entire traffic system simultaneously.

\subsection{Paper Organization}
The remainder of the paper is organized as follows. The problem formulation and system framework of the decision-making issue of CAVs at unsignalized intersections are presented in Section II. In Section III, the algorithm for driving risk assessment is designed. Then, the fuzzy coalitional game approach is applied to the cooperative decision-making issue of CAVs at unsignalized intersections in Section IV. The testing, validation and analysis of the proposed decision-making algorithm are presented in Section V. Finally, Section VI details the conclusion and future work of this paper.

\section{Problem Formulation and System Framework}
\subsection{Decision Making at Unsignalized Intersections}

The unsignalized intersection scenario is illustrated in Fig. 1 (a). The unsignalized intersection, i.e., the conflict zone (CZ), is linked with eight two-lane roads, i.e., $M1$, $M2$, $M3$, $M4$, $\hat{M}1$, $\hat{M}2$, $\hat{M}3$, and $\hat{M}4$.
In the decision-making zone, all vehicles are classified into three types according to the absolute position. The red vehicle and blue vehicles are ready to enter the intersection, which are called ready vehicles (RVs). Green vehicles pass the intersection, which are called passing vehicles (PVs). In addition, black vehicles have moved out of the intersection, which are called the outside vehicles (OVs). The decision-making issue of CAVs at unsignalized intersections is only related to RVs and PVs. The above naming and classification conventions are for managing the entire traffic system at an unsignalized intersection. From the perspective of individual decision making, all vehicles can be divided into four kinds according to the relative relationship and driving conflict, i.e., host vehicle (HV), leading vehicle (LV), neighbor conflict vehicle (NV), and irrelevant vehicle (IV). According to the relative positions of surrounding vehicles, the vehicle driving ahead of the HV is defined as the LV. The vehicle that has a direct driving conflict with the HV is the NV of the HV. Other vehicles are IVs of HV. The effect of the LV on the HV is unidirectional. In this paper, we assume that the decision-making result of HVs will not lead to a change in the LV. NVs are opponents for the HV, and decision-making behaviors are interactive. Considering the large distance between the IV and HV, the effects of driving behaviors are neglected. It should be noted that the names of NV, LV and IV are not fixed. With the change in an HV's motion position, the roles of surrounding vehicles will also change.

\begin{figure}[t]\centering
	\includegraphics[width=7cm]{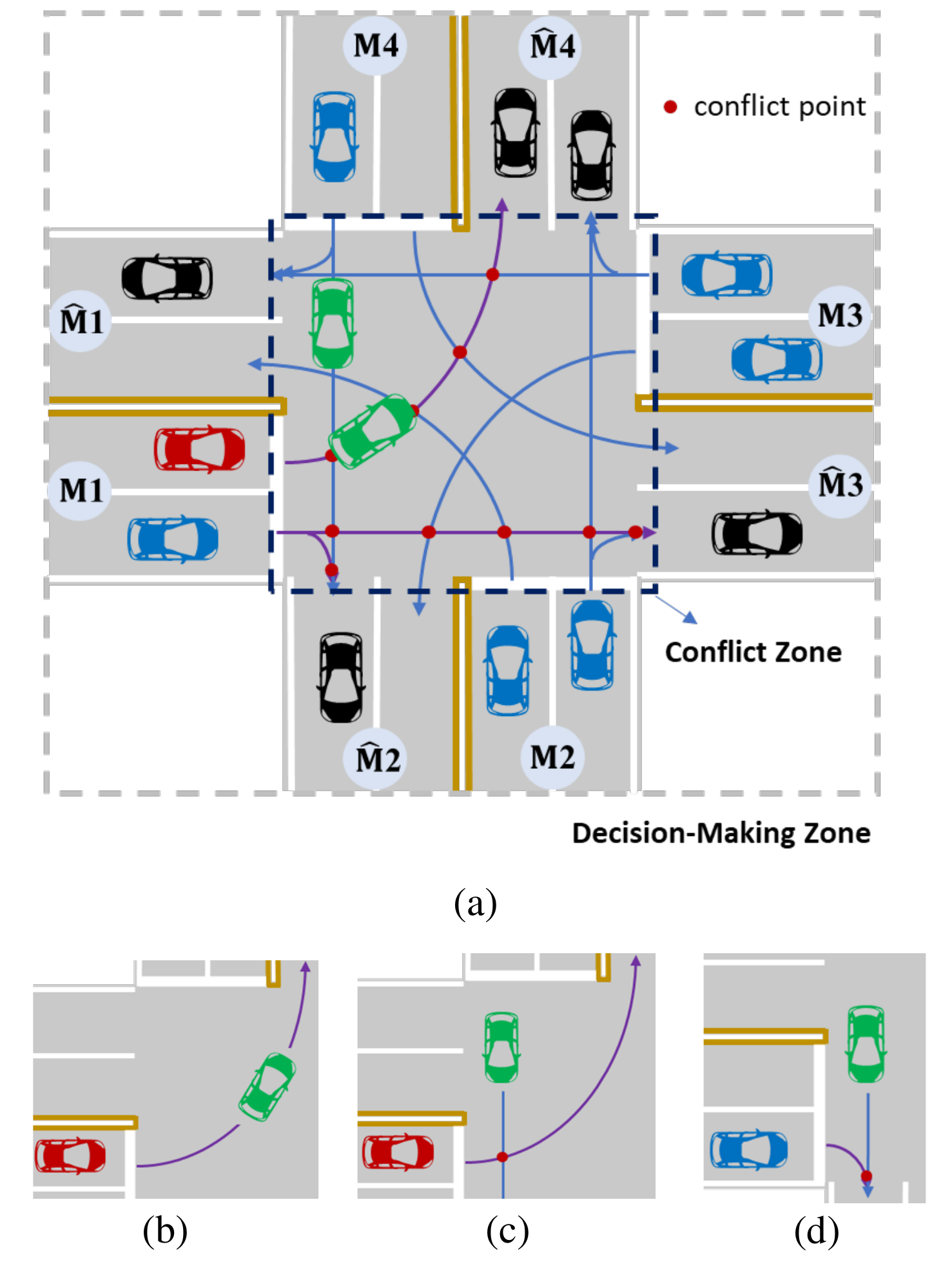}
\caption{Decision-making issue for CAVs at an unsignalized intersection: (a) Conflict network; (b) Following conflict; (c) Cross conflict; (d) Confluence conflict.}\label{FIG_2}
\end{figure}

In the decision-making issue of unsignalized intersections, there are different driving conflicts. Fig. 1 (a) shows the conflict points (CPs) and the distribution network at an unsignalized intersection. All the driving conflicts are divided into three types, i.e., following conflict, cross conflict and confluence conflict, which are displayed in Fig. 1 (b), (c) and (d), respectively. For instance, if the red vehicle wants to turn left from road $M1$ and then merges onto road $\hat{M}4$, there will be four cross CPs and one following conflict on its moving path. Similarly, for the blue vehicle, if it wants to go straight, there will be four cross CPs, one confluence CP and one following conflict. The proposed decision-making algorithm aims to address the driving conflict of CAVs at each CP. Additionally, the effects of the personalized driving characteristics of CAVs on the decision-making results will be studied.

Some assumptions are listed as follows. All vehicles are assumed to be CAVs in a connected driving environment. The information of motion states and positions can be shared with each other for decision making. To realize personalized and human-like driving, the factor of driving aggressiveness is considered in the decision-making model. Besides, all CAVs move according to the traffic network in Fig. 1 (a), the lane-change behavior is not allowed at the conflict zone.

\subsection{Decision-Making Framework for CAVs}
To address the cooperative decision-making issue of CAVs at unsignalized intersections, a game-theoretic decision-making framework is designed in this paper, which is displayed in Fig. 2. In this study, personalized driving characteristics of CAVs are considered in the decision-making issue, which reflect driving aggressiveness. With the increase in driving aggressiveness, CAV gives higher priority to passing efficiency. In contrast, with a decrease in driving aggressiveness, a CAV becomes more conservative and places more emphasis on driving safety. In this paper, most CAVs, analogous to most human drivers, are assumed to possess moderate driving characteristics. Driving aggressiveness is denoted by a variable $\kappa$, which has a significant effect on decision making.

In the game-theoretic decision-making framework, the simplified vehicle kinematic model is constructed for the algorithm design of the driving risk assessment and the decision-making cost function. The driving risk assessment algorithm is proposed to evaluate the safety of the decision-making algorithm and to reduce the amount of calculation of the decision-making cost function. Two key performances for decision making, driving safety and passing efficiency are considered in the cost function. Moreover, multiple constraints are taken into account, including safety, efficiency, comfort, stability and control. Based on the decision-making cost functions with multiple constraints of all CAVs, the fuzzy coalitional game approach is utilized to address the cooperative decision-making issue of CAVs at unsignalized intersections, which can comprehensively take into consideration the social benefit of the whole traffic system and the individual benefit of each personalized CAV. After solving the formulated optimization problem of the game, the decision-making results are sent to the motion planning module of each CAV. Finally, the motion control module conducts the driving behaviors generated from the decision making. In this paper, we mainly focus on the algorithm design of the decision-making module. The algorithms of motion planning and control are provided in previous work \cite{hang2020integrated}.

\begin{figure}[t]\centering
	\includegraphics[width=7cm]{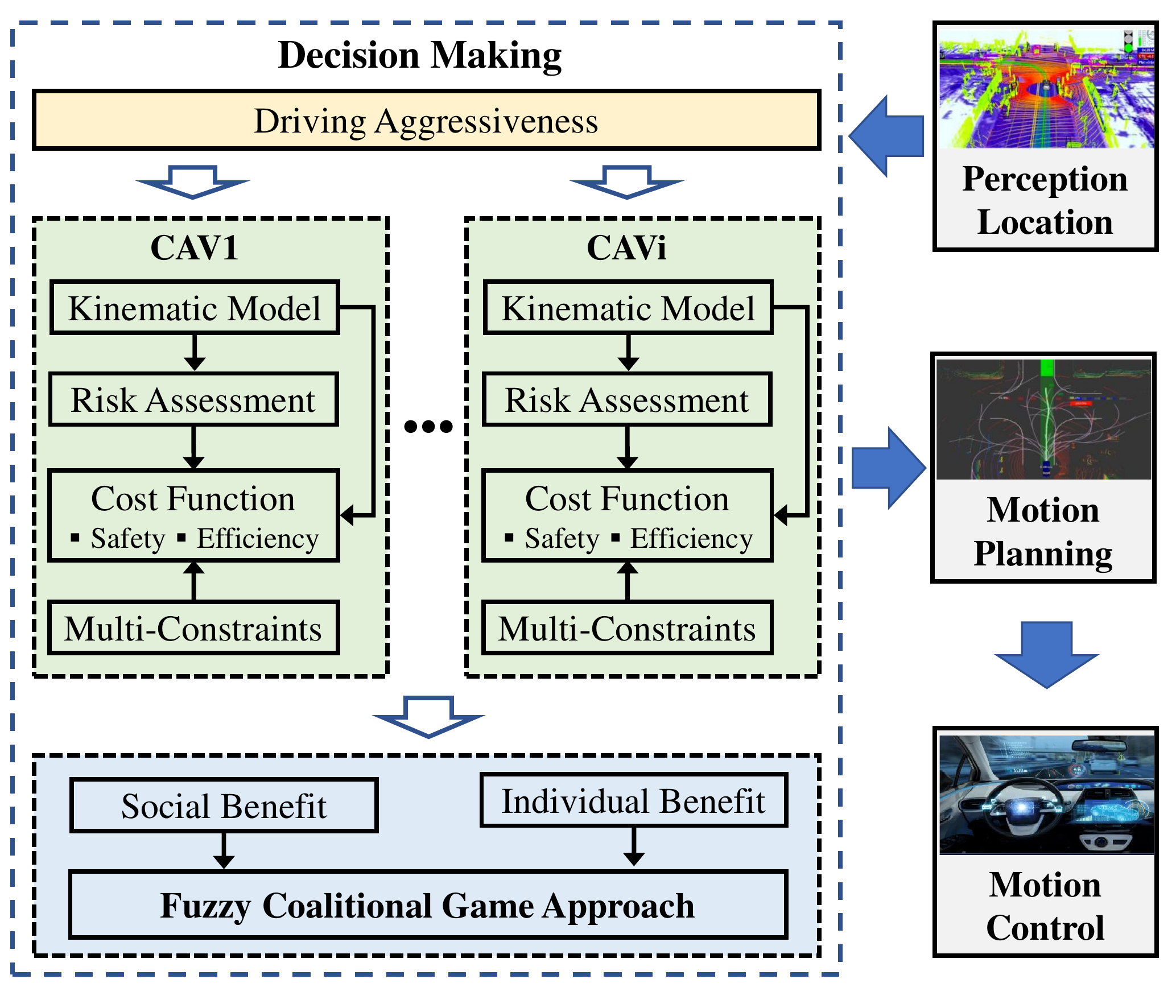}
\caption{Decision-making framework for CAVs.}\label{FIG_3}
\end{figure}

\section{Driving Risk Assessment}
In this section, a simplified single-track vehicle kinematic model is applied to the driving risk assessment of CAVs. In the algorithm design of driving risk assessment, a Gaussian potential field approach is utilized, and some numerical simulations are carried out to verify the effectiveness and feasibility of the algorithm.
\subsection{Vehicle Kinematic Model}
Since vehicles usually pass the interaction with middle or even low speeds and to reduce the computational complexity of the decision-making algorithm, the simplified single-track vehicle kinematic model is adopted, which is derived as follows \cite{hang2021path}:
\begin{align}
\dot{x}(t)=f[x(t),u(t)]
\tag{1a}
\end{align}
\begin{align}
f[x(t),u(t)]=
&
\left[
\begin{array}{ccc}
a_x\\
v_x\tan\beta/l_r\\
v_x\cos(\varphi+\beta)/\cos\beta\\
v_x\sin(\varphi+\beta)/\cos\beta\\
\end{array}
\right]
\tag{1b}
\end{align}
\begin{align}
\beta=\arctan[l_r/(l_f+l_r)\tan\delta_f]
\tag{1c}
\end{align}
where the state vector $x=[v_x, \varphi, X_g, Y_g]^{T}$ and the control vector $u=[a_x, \delta_f]^{T}$. $v_x$ and $\varphi$ are the longitudinal velocity and yaw angle, respectively. $(X_g, Y_g)$ is the coordinate position of the center of gravity. $a_x$ and $\delta_f$ are the longitudinal acceleration and the steering angle of the front wheel, respectively. $\beta$ is the sideslip angle. $l_f$ and $l_r$ are the front and rear wheelbases, respectively.

\subsection{Driving Risk Assessment with Potential Field Approach}
In Subsection A, the four-wheel vehicle model is simplified as a single-track model. For driving risk assessment, Fig. 3 is presented. Point $G$ denotes the center of gravity. Points $G_f$ and $G_r$ denote the centers of the front and rear axles, respectively. Additionally, points $P$ and $C$ denote the predicted point and the turning center, respectively. Path prediction was conducted with model predictive control (MPC), which was discussed in a previous publication \cite{hang2020human}.

The turning radius $R_r$ can be expressed as $1/\rho_r$, where $\rho_r$ is the curvature of the path at point $G_r$.
\begin{align}
\rho_r=\tan\delta_f/l
\tag{2}
\end{align}
where $l$ denotes the wheelbase of the vehicle.

Additionally, the position coordinates of points $G_r$ and $C$ can be derived as follows.
\begin{align}
\begin{array}{ccc}
X_r=X_g-l_r\cos\varphi\\
Y_r=Y_g-l_r\sin\varphi\\
\end{array}
\tag{3a}
\end{align}
\begin{align}
\begin{array}{ccc}
X_c=X_g-l_r\cos\varphi+\sin\varphi/\rho_r\\
Y_c=Y_g-l_r\sin\varphi-\cos\varphi/\rho_r\\
\end{array}
\tag{3b}
\end{align}

\begin{figure}[t]\centering
	\includegraphics[width=5.5cm]{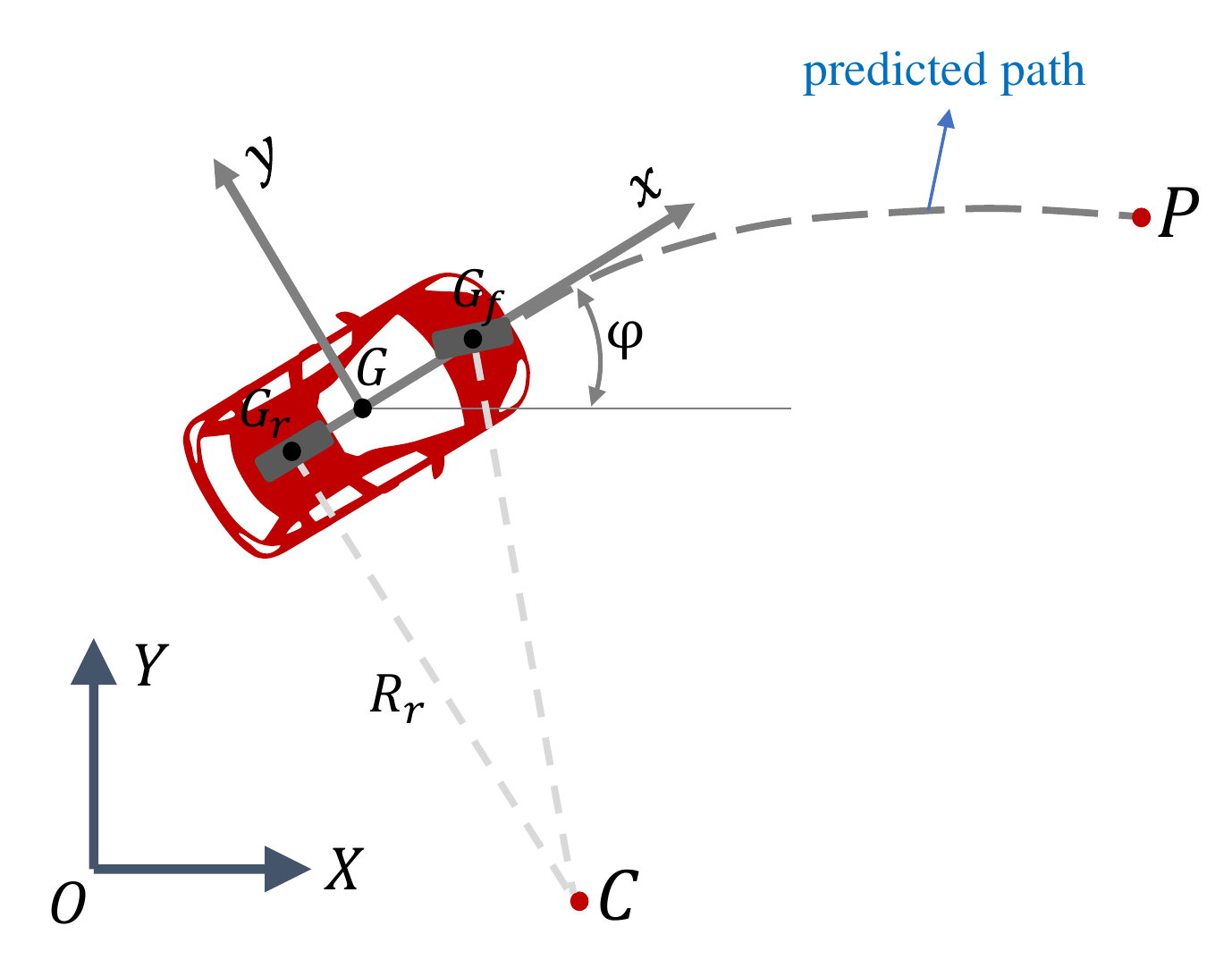}
\caption{Path prediction for risk assessment.}\label{FIG_4}
\end{figure}

\begin{figure}[t]\centering
	\includegraphics[width=8.5cm]{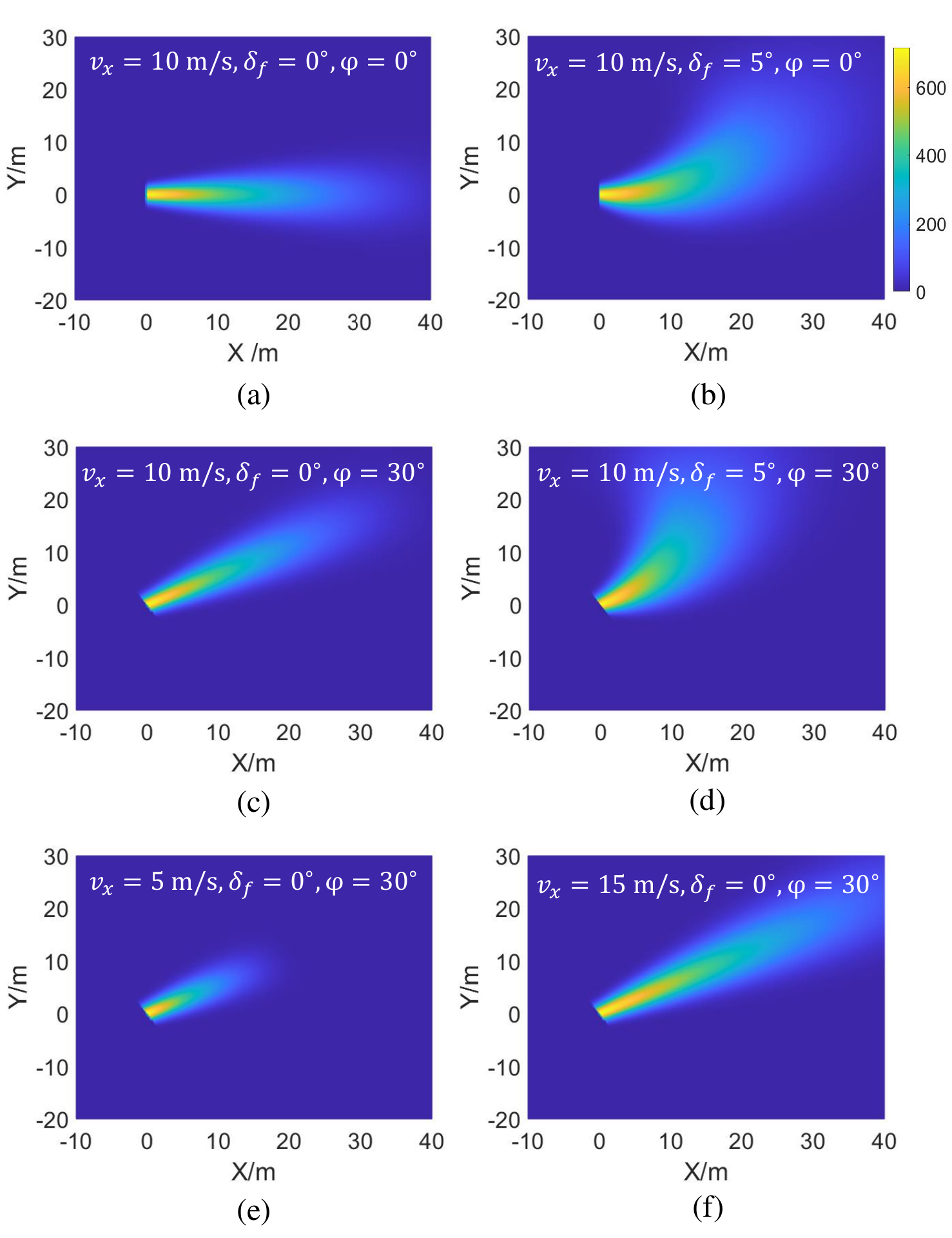}
\caption{Potential field model for risk assessment.}\label{FIG_5}
\end{figure}

Furthermore, the potential field model for driving risk assessment is expressed with a Gaussian cross-section \cite{kolekar2020human}.
\begin{align}
\Gamma=\Lambda\exp[\frac{-(\sqrt{(X-X_c)^2+(Y-Y_c)^2}-1/\rho_r)^2}{2\sigma^2}]
\tag{4}
\end{align}
where the height of Gaussian $\Lambda$ is expressed with a parabolic function, which is related to the length of the predicated path and vehicle velocity.
\begin{align}
\Lambda=a(s-v_xt_p)^2
\tag{5}
\end{align}
where the coefficient $a=a_0 e^\kappa$, $a_0$ is a fixed value, and $\kappa$ denotes the aggressiveness level of driving with $\kappa\in[-1, 1]$. Moreover, $s$ denotes the length of the predicted path, and $t_p$ denotes the predicted time.

The width of the Gaussian distribution, i.e., the standard deviation $\sigma$, is expressed as
\begin{align}
\sigma=(b+c|\delta_f|)s+d
\tag{6}
\end{align}
where $b$ defines the slope of widening of the potential field when $\delta_f=0$, $c$ is the gain coefficient of $|\delta_f|$, $d=W/4$, and $W$ is the width of the vehicle.

To evaluate the effect of key variables on the potential field distribution, a numerical simulation is carried out. The results are illustrated in Fig. 4, indicating that the potential field distribution changes with the variation of $v_x$, $\delta_f$ and $\varphi$. In Fig. 4, the potential field is distributed along the driving direction. The closer to the vehicle, the greater the potential field value is, which means a higher safety risk exists. The change in $\delta_f$ leads to the turning distribution of the potential field along the predicted path. This indicates that the proposed driving risk assessment approach can be used not only for straight driving scenarios but also for turning and curved driving scenarios. Additionally, with the increase of $v_x$, the potential field value and distribution area increase, which means that a higher velocity introduces a greater driving risk.

\section{Decision-Making Algorithm with the Game Theoretic Approach}
In this section, the decision-making cost function is constructed for CAVs considering driving safety and passing efficiency. Then, multiple constraints are defined for the decision-making cost function, including the performances of safety, efficiency, comfort, stability and control. Based on the proposed decision-making cost function and multiple constraints, the fuzzy coalitional game approach is applied to the cooperative decision-making issue for CAVs at unsignalized intersections.

\subsection{Fuzzy Coalitional Game}
The fuzzy coalitional game belongs to the class of cooperative games and aims to minimize the coalition's cost via cooperation. The definition of the fuzzy coalitional game is described as follows.

\textbf{Definition 1} \cite{chen2010stimulating}: In a game issue, there exist $n$ players, and the set of all players is denoted by $N=\{1,2,\cdots,n\}$. Each subset $S$ of $N$ is called a coalition, i.e., $S\in2^N$. $S$ can be an empty set, i.e., $S=\emptyset$. If $S$ consists of only one player, it is called a single-player coalition. If $S$ consists of all players, i.e., $S=N$, it is called a grand coalition. The coalitional game model is represented by a pair $\langle N,U,V\rangle$, where $U$ denotes the set of decision-making behaviors of players, and $V$ denotes the characteristic function, $V(\emptyset)=0$. In a typical game issue, the characteristic function $V$ is usually denoted by a payoff function. Each coalition aims to maximize the payoff value. However, regarding the decision-making issue of CAVs at the unsignalized intersection in this paper, $V$ corresponds to the minimum cost. The construction of the cost function is presented in the next subsection.

The above definition provides a description of the conventional coalitional game. However, it is based on the assumption that each player can only join one coalition. Actually, players may choose not to limit contributing their sources to only one coalition. In particular, distributed contributions lead to the concept of a fuzzy coalition game, in which the players can allocate their sources and join several coalitions. The participation coefficient $p_i^S$ is used to describe the source allocation of player $i$ in the fuzzy coalition $S$. As a result, the participation coefficients of all the players in the fuzzy coalition $S$ yield the following vector:
\begin{align}
p^S=[p_1^S, p_2^S, \cdots, p_i^S, \cdots, p_n^S]^T, \quad p_i^S\in[0, 1]
\tag{7}
\end{align}

If all participation coefficients in the fuzzy coalition $S$ equal 0 or 1, the fuzzy coalitional game issue degenerates to a conventional coalitional game. Specifically, the conventional coalitional game is simply a special case of the fuzzy coalitional game.

Regarding the cost allocation of the fuzzy coalitional game, different approaches or algorithms can be used, e.g., the fuzzy minimum kernel method and fuzzy Shapley method \cite{meng2015induced}. Then, the allocated cost of all players in the fuzzy coalition $S$ forms the vector $H^S=(H_1^S, H_2^S, \cdots, H_i^S, \cdots, H_n^S)$. If $H^S$ satisfies individual rationality, collective rationality and superadditivity, $H^S$ can be regarded as the solution to the fuzzy coalitional game issue \cite{bui2018cooperative}.

Individual rationality requires that the player in the fuzzy coalition obtains a satisfactory cost that is no more than that incurred by running separately with the same engagement, which is described as follows.
\begin{align}
H_i^S\leq V(e^i\cdot p_i^S)
\tag{8}
\end{align}
where vector $e^i=[0,0,\cdots,1,\cdots,0]^T$, in which the $ith$ element is 1 and other elements are 0.

Collective rationality requires that the cost of the fuzzy coalition be allocated to all players at once.
\begin{align}
V(p^S)=\sum_{i\in S}H_i^S
\tag{9}
\end{align}

Superadditivity means that the cost of any fuzzy coalition is no more than the total cost of all players in the fuzzy coalition running separately with the same engagement.
\begin{align}
V(p^S)\leq \sum_{i\in S} V(e^i\cdot p_i^S)
\tag{10}
\end{align}

In this paper, the decision-making issue of CAVs at unsignalized intersections is regarded as a fuzzy coalitional game. All RVs and PVs are the game players. Once the CAV changes from a PV to an OV, it will not be a game player. Both the social benefit of the entire traffic system and the individual benefit of each personalized CAV are considered in the decision-making issue. Two types of coalitions, i.e., the grand coalition and single-player coalition, are proposed to construct the fuzzy coalition. Grand coalition $S_g$ includes all game players, which represents the entire traffic system. Single-player coalition $S_i$ reflects the individual benefit the CAV, namely, CAV$_i$.
 Furthermore, the participation vector of all the players in the fuzzy coalition $S_g$ is expressed as
\begin{align}
p^{S_g}=[p_1^{S_g}, p_2^{S_g}, \ldots, p_i^{S_g}, \ldots, p_n^{S_g}]^T, \quad p_i^{S_g}\in[0, 1]
\tag{11}
\end{align}

The participation vector of CAV$_i$ in the fuzzy coalition $S_i$ is expressed as $p_i^{S_i}$. CAV$_i$ only joins two fuzzy coalitions, i.e., $S_g$ and $S_i$. It yields that $p_i^{S_g}+p_i^{S_i}=1$. In this paper, participation $p_i^{S_g}$ is assumed to be related to driving aggressiveness $\kappa$, which obeys a Gaussian distribution.
\begin{align}
p_i^{S_g}=\frac{1}{\sqrt{2\pi}\hat{\sigma}}\exp[\frac{-(\kappa^i-\hat{\mu})^2}{2\hat{\sigma}^2}]
\tag{12}
\end{align}
where $\hat{\mu}=0$, $\hat{\sigma}=1/\sqrt{2\pi}$, and $\kappa^i$ denotes the driving aggressiveness of CAV$_i$.

\subsection{Cost Function Construction for Decision Making}
As described in Section II, the decision-making issue of CAVs at the intersection addresses the driving conflicts at the CZ. In particular, the decision-making algorithm should ensure the safety of each CAV at each CP. In addition, the passing efficiency of the entire traffic system is another vital factor. Both safety and efficiency are critical for CAVs in decision making, which are conflicting at times. The two critical performances are associated with the decision-making results of CAVs, which are reflected by the driving behaviors, i.e., accelerating, decelerating and steering. The accelerating or decelerating control is related to the control vector $a_x$, and the steering control is decided by the control vector $\delta_f$. Therefore, $a_x$ and $\delta_f$ are two decision-making variables in the decision-making cost function. With the control of decision-making variables and optimization of the decision-making cost function, CAVs can obtain good driving performances at unsignalized intersections.

In the decision-making cost function construction, the kinds of driving performance, i.e., driving safety and passing efficiency, are considered. For HV (CAV$_i$), the cost function for decision making consists of the cost of driving safety $V_s^{i}$, and the cost of passing efficiency $V_e^{i}$.
\begin{align}
V^{i}=k_s^{i}V_s^{i}+k_e^{i}V_e^{i}
\tag{13a}
\end{align}
\begin{align}
k_s^{i}=\frac{e^{-\kappa+1}}{e^{-\kappa+1}+e^{\kappa+1}},\quad k_e^{i}=\frac{e^{\kappa+1}}{e^{-\kappa+1}+e^{\kappa+1}}
\tag{13b}
\end{align}
where $k_s^{i}$ and $k_e^{i}$ are the weighting coefficients, which vary with the change of $\kappa$. Furthermore, with the increase in the aggressiveness level, the cost of passing efficiency has a larger weighting coefficient. Otherwise, $V^{i}$ is more concerned with driving safety.

The cost function of driving safety $V_s^{i}$ consists of three cost functions, i.e., the cost functions of longitudinal, lateral and lane-keeping safety, denoted by $V_{s-log}^{i}$, $V_{s-lat}^{i}$ and $V_{s-lk}^{i}$, respectively.
\begin{align}
V_s^{i}=\omega_{s\sim log}^{i} V_{s\sim log}^{i}+\omega_{s\sim lat}^{i} V_{s\sim lat}^{i}+\omega_{s\sim lk}^{i} V_{s\sim lk}^{i}
\tag{14}
\end{align}
where $\omega_{s\sim log}^{i}$, $\omega_{s\sim lat}^{i}$ and $\omega_{s\sim lk}^{i}$ are the weighting coefficients. The values of $\omega_{s\sim log}^{i}$ and $\omega_{s\sim lat}^{i}$ are related to the driving risk assessment, $\omega_{s\sim lk}^{i}=1$.

The cost function of longitudinal safety $V_{s\sim log}^{i}$ is defined with the time to collision (TTC) between CAV$_i$ and its LV, which mainly addresses the following conflict.
\begin{align}
V_{s\sim log}^{i}=(k_{log}^{i}/TTC_{log}^{i})^2
\tag{15a}
\end{align}
\begin{align}
TTC_{log}^{i}=\Delta s^{i\sim LV}/\Delta v_{x}^{i\sim LV}
\tag{15b}
\end{align}
\begin{align}
\Delta v_{x}^{i\sim LV}=v_{x}^{i}-v_{x}^{LV}
\tag{15c}
\end{align}
\begin{align}
k_{log}^{i}=
&
\left\{
\begin{array}{lr}
1,\quad \Delta v_{x}^{i\sim LV}\geq0\\
0,\quad \Delta v_{x}^{i\sim LV}<0\\
\end{array}
\right.
\tag{15d}
\end{align}
where $v_{x}^{LV}$ and $v_{x}^{i}$ denote the longitudinal velocities of LV and CAV$_i$, respectively. $\Delta v_{x}^{i\sim LV}$ and $\Delta s_{x}^{i\sim LV}$ denote the relative velocity and longitudinal gap between CAV$_i$ and LV. According to the definition of $k_{log}^{i}$, if the velocity of CAV$_i$ is smaller than that of LV, the longitudinal safety cost will be ignored.

The cost function of lateral safety $V_{s\sim lat}^{i}$ is defined with the time to collision (TTC) between CAV$_i$ and its NV at the CP, which mainly addresses cross and confluence conflicts.
Assuming that there exist $m$ CPs on the passing path of CAV$_i$, at the $jth$ CP, $j\in\{1,2,\cdots,m\}$, the lateral safety cost $V_{s\sim lat}^{i}$ is expressed as
\begin{align}
V_{s\sim lat}^{i}=e^j\cdot V_{s\sim lat}^{i\sim CP}
\tag{16a}
\end{align}
\begin{align}
V_{s\sim lat}^{i\sim CP}=[V_{s\sim lat}^{i\sim CP1}, V_{s\sim lat}^{i\sim CP2}, \cdots , V_{s\sim lat}^{i\sim CPj},\cdots,V_{s\sim lat}^{i\sim CPm}]^T
\tag{16b}
\end{align}
\begin{align}
V_{s\sim lat}^{i\sim CPj}=[(TTC_{lat}^{i}-TTC_{lat}^{NVj})^2+\xi]^{-1}
\tag{16c}
\end{align}
\begin{align}
TTC_{lat}^{i}=\Delta s^{i\sim CPj}/v_{x}^{i}
\tag{16d}
\end{align}
\begin{align}
TTC_{lat}^{NVj}=\Delta s^{NVj\sim CPj}/v_{x}^{NVj}
\tag{16e}
\end{align}
where vector $e^j=[0,0,\cdots,1,\cdots,0]^T_{m\times 1}$, in which the $jth$ element is 1 and other elements are 0. In particular, if the safety risk values at the ahead CPs of the $jth$ CP are beyond the safe set, $e^j$ will be changed. $V_{s\sim lat}^{i\sim CP}$ denotes the set of lateral safety costs at all CPs. At the $jth$ CP, the lateral safety cost $V_{s\sim lat}^{i\sim CPj}$ is described by Eq. 16c, in which $TTC_{lat}^{i}$ and $TTC_{lat}^{NVj}$ denote the TTCs of CAV$_i$ and NV$_j$ to the $jth$ CP, respectively. $\Delta s^{i\sim CPj}$ and $\Delta s^{NVj\sim CPj}$ denote the distances between CAV$_i$ and the $jth$ CP and between NV$_j$ and the $jth$ CP, respectively. $v_{x}^{NVj}$ is the longitudinal velocity of NV$_j$. Additionally, $\xi$ is a design parameter consisting of a small positive value to avoid a zero denominator in the calculation.

The third part of $V_s^i$ is the cost of lane-keeping safety $V_{s\sim lk}^{i}$, defined with the lateral distance error $\Delta y^{i}$ and the yaw angle error $\Delta \varphi^{i}$ between the position of CAV$_i$ and the target path, which is derived as
\begin{align}
V_{s\sim lk}^{i}=k_{y\sim lk}^{i}(\Delta y^{i})^2+k_{\varphi\sim lk}^{i}(\Delta \varphi^{i})^2
\tag{17}
\end{align}
where $k_{y\sim lk}^{i}$ and $k_{\varphi\sim lk}^{i}$ are the weighting coefficients, $k_{y\sim lk}^{i}=1$, $k_{\varphi\sim lk}^{i}=80$.

Finally, the cost function of passing efficiency $V_{e}^{i}$ for CAV$_i$ is defined with the time headway $THW^i$ of CAV$_i$, which is expressed as
\begin{align}
V_{e}^{i}=(THW^{i})^2
\tag{18a}
\end{align}
\begin{align}
THW^{i}=\Delta s^{i\sim LV}/v_{x}^{i}
\tag{18b}
\end{align}

To reduce the computation of the decision-making algorithm, the weighting coefficients $\omega_{s\sim log}^{i}$ and $\omega_{s\sim lat}^{i}$ in Eq. 14 are defined according to the driving risk assessment.
\begin{align}
\omega_{s\sim log}^{i}=
&
\left\{
\begin{array}{lr}
\omega_0,\quad\Gamma^i(LV)>\Gamma_0\\
0,\quad\quad\Gamma^i(LV)\leq\Gamma_0\\
\end{array}
\right.
\tag{19a}
\end{align}
\begin{align}
\omega_{s\sim lat}^{i}=
&
\left\{
\begin{array}{lr}
\omega_0,\quad\Gamma^{i}(NVj)>\Gamma_0 \quad \mathrm{or} \quad \Gamma^{NVj}(i)>\Gamma_0\\
0,\quad\quad \mathrm{otherwise}\\
\end{array}
\right.
\tag{19b}
\end{align}
where $\Gamma_0$ denotes the safe value of driving risk assessment, and $\omega_0$ denotes a fixed weighting coefficient, $\omega_0=10$. Besides, $\Gamma^{i}(NVj)$ denotes the driving risk assessment value from NV$_j$ for CAV$_i$, and $\Gamma^{NVj}(i)$ denotes the driving risk assessment value from CAV$_i$ for NV$_j$. Either $\Gamma^{i}(NVj)$ or $\Gamma^{NVj}(i)$  is larger than the safe value, the cost function of lateral driving safety should be considered.

\subsection{Constraints of Decision Making}
To ensure the performance of the decision-making algorithm, including driving safety, passing efficiency, riding comfort, the control vector, and lateral stability, some constraints must be taken into account. The safety constraints for CAV$_i$ are expressed as
\begin{align}
TTC^{i}\geq TTC^{\mathrm{min}},|\Delta y^{i}|\leq\Delta y^{\mathrm{max}},|\Delta \varphi^{i}|\leq\Delta \varphi^{\mathrm{max}}
\tag{20}
\end{align}

Moreover, the constraint for travel efficiency is defined by
\begin{align}
v_{x}^{i}\leq v_{x}^{\mathrm{max}}
\tag{21}
\end{align}

The constraint for riding comfort is related to the jerk, which is expressed as
\begin{align}
|jerk^{i}|\leq jerk^{\mathrm{max}}
\tag{22}
\end{align}

Additionally, the control constraints of the control vectors $a_x^{i}$ and $\delta_f^{i}$ are given by
\begin{align}
|a_x^{i}|\leq a_x^{\mathrm{max}},\quad |\delta_f^{i}|\leq \delta_f^{\mathrm{max}}
\tag{23}
\end{align}

The lateral stability constraint is associated with the sideslip angle [5], which is given by
\begin{align}
|\beta^{i}|\leq \arctan(0.02\mu g)
\tag{24}
\end{align}
where $\mu$ denotes the tire-road adhesion coefficient.

Finally, the aforementioned constraints for CAV$_i$ can be expressed in a compact form as
\begin{align}
\begin{array}{lr}
\Pi^{i}(TTC^{i}, \Delta y^{i},\Delta \varphi^{i}, v_{x}^{i}, jerk^{i}, a_{x}^{i},\delta_f^{i},\beta^{i})\leq0
\tag{25}
\end{array}
\end{align}

The constraint boundaries for decision making are presented in Table I.
\begin{table}[h]
	\renewcommand{\arraystretch}{1.5}
	\caption{Constraint Boundaries for Decision Making}
\setlength{\tabcolsep}{6mm}
	\centering
	\label{table_1}
	\resizebox{\columnwidth}{!}{
		\begin{tabular}{c c | c c}
			\hline\hline \\[-4mm]
            Parameter & Value & Parameter & Value \\
\hline
			$TTC^{\mathrm{min}}$/ $\mathrm{(s)}$  & 1.5 & $v_{x}^{\mathrm{max}}$/ $\mathrm{(m/s)}$ & 8\\
           $\Delta y^{\mathrm{max}}$/ $\mathrm{(m)}$  & 0.2 & $a_x^{\mathrm{max}}$/ $\mathrm{(m/s^2)}$ & 8 \\
           $\Delta \varphi^{\mathrm{max}}$/ $\mathrm{(deg)}$  & 2 & $\delta_f^{\mathrm{max}} $/ $\mathrm{(deg)}$ & 30\\
           $jerk^{\mathrm{max}}$/ $\mathrm{(m/s^3)}$ & 2 & -- & --\\
			\hline\hline
		\end{tabular}
	}
\end{table}

\subsection{Decision Making with the Fuzzy Coalitional Game Approach}
In this paper, two kinds of fuzzy coalitions, $S_g$ and $S_i, i\in N=\{1,2,\cdots,n\}$, are constructed for decision making. In the fuzzy coalition $S_g$, all players, i.e., RVs and PVs, aim to maximize the benefit of $S_g$, i.e., the entire traffic system. In the fuzzy coalition $S_i$, CAV$_i$ tries to pursue the maximum individual benefit. Based on the above analysis, CAV$_i$ must consider both self-interest and collective interest in the decision-making process. The weights are reflected by the participation vector described in Eqs. (11) and (12). The decision-making vector of CAV$_i$ is defined as $u^{i}=[ a_x^{i},\delta_f^{i}]^T$.
The decision-making vector of all players is written as
\begin{align}
U=[u^1, u^2, \cdots, u^i, \cdots, u^n]^T
\tag{26}
\end{align}

The decision-making cost function vector of all players is expressed as
\begin{align}
\emph{\textbf{V}}=[V^1, V^2, \cdots, V^i, \cdots, V^n]^T
\tag{27}
\end{align}

Furthermore, the decision-making cost function for the fuzzy coalition $S_g$ can be derived as
\begin{align}
V^{S_g}=p^{S_g}\cdot\emph{\textbf{V}}
\tag{28}
\end{align}

The decision-making cost function for the fuzzy coalition $S_i$ is expressed as
\begin{align}
V^{S_i}=p_i^{S_i}\cdot V^i
\tag{29}
\end{align}

Based on the established decision-making cost functions and constraints, the decision-making issue for CAVs at unsignalized intersections can be transformed into a fuzzy coalitional game, which is expressed as follows.
\begin{align}
\left\{
\begin{array}{lr}
\underset{U}{\min} V^{S_g}\\
\underset{U}{\min} V^{S_1}, \underset{U}{\min} V^{S_2}, \cdots, \underset{U}{\min} V^{S_i}, \cdots, \underset{U}{\min} V^{S_n}\\
U^*=[u^{1*}, u^{2*}, \cdots, u^{i*}, \cdots, u^{n*}]^T\\
\end{array}
\right.
\tag{30}
\end{align}
s.t. $\Pi^i\leq0$, $\forall i\in N$.

Notably, Eq. (30) describes a multilevel optimization issue with constraints. High-level optimization maximizes the social profit for the entire traffic system, and low-level optimization maximizes the individual profit for each CAV. Finally, the decision-making vector $U^*$ can be determined with the active-set optimization algorithm \cite{bard1988convex}. The optimization algorithm updates an estimate of the Hessian of the Lagrangian at each iteration using the BFGS formula. In the fuzzy coalition, the fuzzy Shapley method is used to allocate the benefit for each player, which satisfies
individual rationality, collective rationality and superadditivity. The convergence analysis of the fuzzy coalitional game approach is presented in \cite{aubin1981cooperative}, which will not be introduced repeatedly.

Additionally, we note that with the change of participation vector $p^{S_g}$ reflected by $\kappa$ of each CAV, the optimized decision-making vector will be different. There exist two extreme cases. The first case is $p^{S_g}=\textbf{0}$, which means that all CAVs will not join the fuzzy coalition $S_g$. In particular, each CAV is only concerned about individual benefits, and the social benefit of the entire traffic system is neglected. Hence, the fuzzy coalitional game degenerates to a noncooperative game.
\begin{align}
\left\{
\begin{array}{lr}
\min V^{1}, \min V^{2}, \cdots, \min V^{i}, \cdots, \min V^{n}\\
u^{i*}=\arg\min V^{i}\\
U^*=[u^{1*}, u^{2*}, \cdots, u^{i*}, \cdots, u^{n*}]^T\\
\end{array}
\right.
\tag{31}
\end{align}
s.t. $\Pi^i\leq0$, $\forall i\in N$.

Another extreme case is $p^{S_g}=\textbf{\emph{I}}$, which means that all CAVs will join the fuzzy coalition $S_g$ with $100\%$ participation. Specifically, all CAVs only care about the social benefit of the entire traffic system. Hence, the individual benefit of each CAV is neglected. In this scenario, the fuzzy coalitional game becomes a cooperative global coalitional game, which is expressed as follows.
\begin{align}
U^*=\arg\min \sum\limits_{i=1}^{n} V^{i}
\tag{32}
\end{align}
s.t. $\Pi^i\leq0$, $\forall i\in N$.

In general, for CAV$_i$ (RV and PV) to pass the unsignalized intersection, its decision-making process is summarized in Algorithm 1.

\begin{algorithm}[h]
\caption{Decision making for CAV$_i$ to pass the unsignalized intersection.}
\begin{algorithmic}[1]
\FOR{$i=1:n$}
\STATE Input the motion state and position of CAV$_i$;\
\STATE Calculate the CPs on CAV$_i$'s moving path at the CZ;
\STATE Driving risk assessment to detect LV and NV$_j$;
\FOR{$j=1:m$}
\IF {$\quad\Gamma^i(LV)>\Gamma_0$}
\STATE $\omega_{s\sim log}^{i}=\omega_0$;
\ELSE
\STATE $\omega_{s\sim log}^{i}=0$;
\ENDIF

\IF {$\Gamma^{i}(NVj)>\Gamma_0 \quad \mathrm{or} \quad \Gamma^{NVj}(i)>\Gamma_0$}
\STATE $\omega_{s\sim lat}^{i}=\omega_0$;
\ELSE
\STATE $\omega_{s\sim lat}^{i}=0$;
\ENDIF

\STATE Construct the cost function of CAV$_i$ $V^i$ and the constraint $\Pi^i$;
\STATE Calculate the participation vector $p^{S_g}$;
\STATE Construct the fuzzy coalitions, $S_g$ and $S_i$;
\STATE Solve the fuzzy coalitional game issue, Eq. 26 to Eq. 30;
\STATE Feasibility verification. If there is no solution or Eq. 8 to Eq. 10 are not satisfied, $p_i^{S_g}=0$, back to Step 19;
\STATE Output the decision-making vector $U^*$;
\ENDFOR
\ENDFOR
\end{algorithmic}
\end{algorithm}

\section{Testing, Validation and Analysis}
To evaluate the performance of the game theoretic-based decision-making algorithm for CAVs at unsignalized intersections, three test cases are designed and conducted considering the driving aggressiveness of CAVs. All the test scenarios and algorithm verifications are established and implemented via the MATLAB/Simulink platform in this section.

\begin{table}[t]
	\renewcommand{\arraystretch}{1.3}
	\caption{Aggressiveness Setting for CAVs in Case 1}
\setlength{\tabcolsep}{4.5 mm}
	\centering
	\label{table_3}
	\resizebox{\columnwidth}{!}{
		\begin{tabular}{c c c c c c c }
			\hline\hline \\[-4mm]
			\multirow{2}{*}{Aggressiveness  $\kappa$ } & \multicolumn{6}{c}{Scenario} \\
\cline{2-7} & \makecell [c] {A} & \makecell [c] {B} & \makecell [c] {C} & \makecell [c] {D} & \makecell [c] {E} & \makecell [c] {F} \\
\hline
			\multicolumn{1}{c}{V1} & -0.8 & 0 & 1 & 1 & -0.8 & 1 \\
			\multicolumn{1}{c}{V2} & 0 & 0 & 0 & 0.8 & -0.8 & 0.8 \\
  \multicolumn{1}{c}{V3} & 0 & 0 & 0 & 0 & -0.8 & 1 \\

			\hline\hline
		\end{tabular}
	}
\end{table}

\begin{figure}[!t]\centering
	\includegraphics[width=8.5cm]{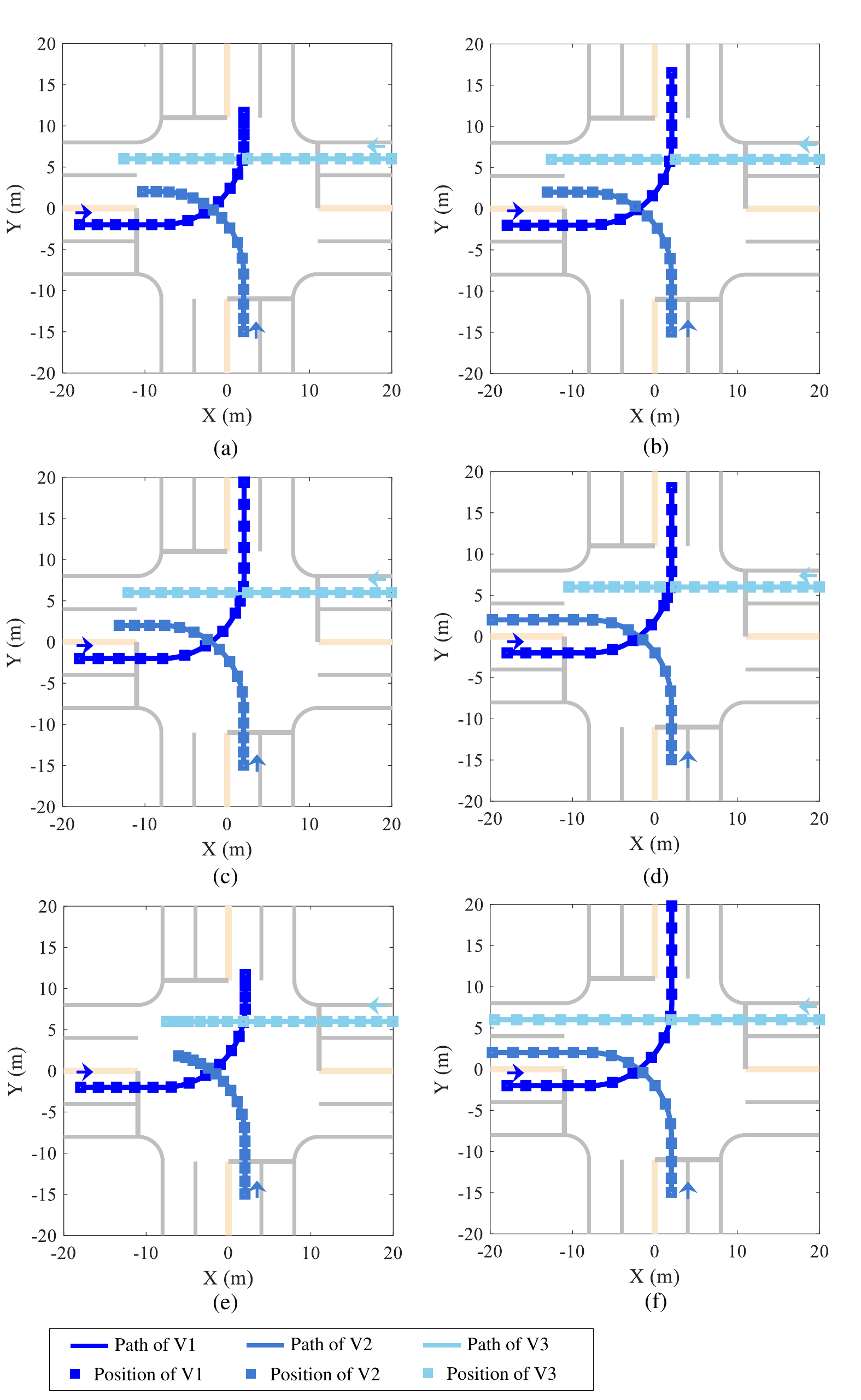}
\caption{Decision-making results of CAVs in Case 1: (a) Scenario A; (b) Scenario B; (c) Scenario C; (d) Scenario D; (e) Scenario E; (f) Scenario F.}\label{FIG_6}
\end{figure}
\subsection{Case 1}

\begin{table*}[!t]
	\renewcommand{\arraystretch}{1.3}
	\caption{Velocities and accelerations of CAVs in Case 1}
\setlength{\tabcolsep}{2 mm}
	\centering
	\label{table_3}
	\resizebox{\textwidth}{19mm}{
		\begin{tabular}{c c c c | c c c | c c c | c c c | c c c | c c c c c c c c c}
			\hline\hline \\[-4mm]
			\multirow{2}{*}{Test results} & \multicolumn{3}{c|}{Scenario A} & \multicolumn{3}{c|}{Scenario B} & \multicolumn{3}{c|}{Scenario C} & \multicolumn{3}{c|}{Scenario D} & \multicolumn{3}{c|}{Scenario E} & \multicolumn{3}{c}{Scenario F} \\
\cline{2-19} & \makecell [c] {V1} & \makecell [c] {V2} & \makecell [c] {V3} & \makecell [c] {V1} & \makecell [c] {V2} & \makecell [c] {V3} & \makecell [c] {V1} & \makecell [c] {V2} & \makecell [c] {V3} & \makecell [c] {V1} & \makecell [c] {V2} & \makecell [c] {V3} & \makecell [c] {V1} & \makecell [c] {V2} & \makecell [c] {V3} & \makecell [c] {V1} & \makecell [c] {V2} & \makecell [c] {V3}\\
\hline
			\multicolumn{1}{l}{Velocity Max / (m/s)} & 5.58 & 5.08 & 5.89 & 6.17 & 5.11 & 5.90 & 6.92 & 5.10 & 5.83 & 7.58 & 6.47 & 5.83 & 5.54 & 4.12 & 5.11 & 7.58 & 6.47 & 6.91\\
			\multicolumn{1}{l}{Velocity RMS / (m/s)} & 4.96 & 4.19 & 5.43 & 5.71 & 4.64 & 5.43 & 6.26 & 4.64 & 5.34 & 6.13 & 5.76 & 5.09 & 4.97 & 3.53 & 4.63 & 6.71 & 5.76 & 6.58\\
\multicolumn{1}{l}{Acceleration Max / $\mathrm{(m/s^2)}$} & 0.89 & 0.53 & 0.35 & 1.13 & 0.53 & 0.35 & 5.64 & 0.53 & 0.76 & 7.67 & 2.22 & 0.53 & 0.21 & 0.06 & 0.04 & 6.40 & 2.23 & 2.54\\
\multicolumn{1}{l}{Acceleration RMS / $\mathrm{(m/s^2)}$}  & 0.25 & 0.32 & 0.20 & 0.17 & 0.28 & 0.20 & 1.42 & 0.28 & 0.21 & 2.17 & 0.78 & 0.26 & 0.02 & 0.03 & 0.02 & 0.97 & 0.78 & 0.64\\
\multicolumn{1}{l}{Jerk Max / $\mathrm{(m/s^3)}$} & 0.64 & 2.00 & 0.04 & 0.29 & 2.00 & 0.05 & 2.00 & 0.01 & 2.00 & 2.00 & 2.00 & 2.00 & 2.00 & 0.01 & 0.15 & 2.00 & 2.00 & 2.00 \\
\multicolumn{1}{l}{Jerk RMS / $\mathrm{(m/s^3)}$} & 0.15 & 0.01 & 0.01 & 0.08 & 0.01 & 0.02 & 0.20 & 0.02 & 0.29 & 0.36 & 0.09 & 0.26 & 0.02 & 0.01 & 0.01 & 0.28 & 0.10 & 0.09 \\
\hline
\multicolumn{1}{l}{System Velocity RMS / (m/s)}  & \multicolumn{3}{c|}{4.89} & \multicolumn{3}{c|}{5.28} & \multicolumn{3}{c|}{5.46} & \multicolumn{3}{c|}{5.68} & \multicolumn{3}{c|}{4.42} & \multicolumn{3}{c}{6.37}\\
			\hline\hline
		\end{tabular}
	}
\end{table*}

\begin{table*}[!t]
	\renewcommand{\arraystretch}{1.2}
	\caption{Relative Distance and TTC in Case 1}
\setlength{\tabcolsep}{3 mm}
	\centering
	\label{table_4}
	\resizebox{\textwidth}{9mm}{
		\begin{tabular}{c c c | c c | c c | c c | c c | c c c c c c}
			\hline\hline \\[-4mm]
			\multirow{2}{*}{Test results} & \multicolumn{2}{c|}{Scenario A} & \multicolumn{2}{c|}{Scenario B} & \multicolumn{2}{c|}{Scenario C} & \multicolumn{2}{c|}{Scenario D} & \multicolumn{2}{c|}{Scenario E} & \multicolumn{2}{c}{Scenario F} \\
\cline{2-13} & \makecell [c] {V1-V2} & \makecell [c] {V1-V3} & \makecell [c] {V1-V2} & \makecell [c] {V1-V3} & \makecell [c] {V1-V2} & \makecell [c] {V1-V3} & \makecell [c] {V1-V2} & \makecell [c] {V1-V3} & \makecell [c] {V1-V2} & \makecell [c] {V1-V3} & \makecell [c] {V1-V2} & \makecell [c] {V1-V3} \\
\hline
			\multicolumn{1}{c}{Distance Min / (m)} & 6.00 & 6.25 & 5.35 & 5.50 & 6.79 & 4.45 & 4.64 & 4.39 & 6.35 & 6.29 & 4.64 & 5.85 \\
			\multicolumn{1}{c}{TTC Min / (s)} & 2.16 & 2.07 & 2.01 & 2.06 & 2.33 & 1.60 & 1.72 & 1.54 & 2.41 & 2.07 & 1.72 & 2.04 \\

			\hline\hline
		\end{tabular}
	}
\end{table*}

\begin{figure}[!t]\centering
	\includegraphics[width=8.5cm]{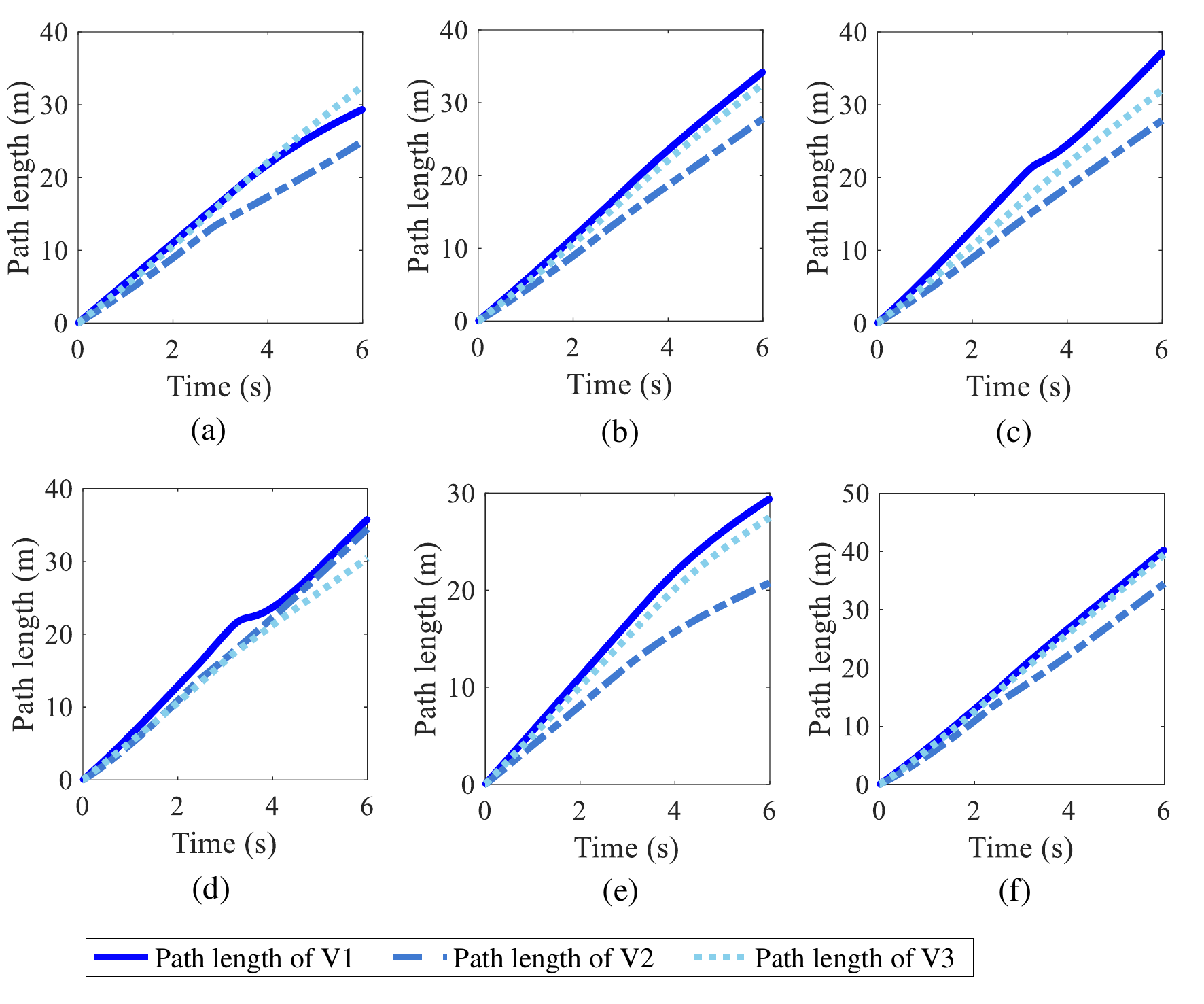}
\caption{Path length of CAVs in Case 1: (a) Scenario A; (b) Scenario B; (c) Scenario C; (d) Scenario D; (e) Scenario E; (f) Scenario F.}\label{FIG_7}
\end{figure}

\begin{figure}[!t]\centering
	\includegraphics[width=8.5cm]{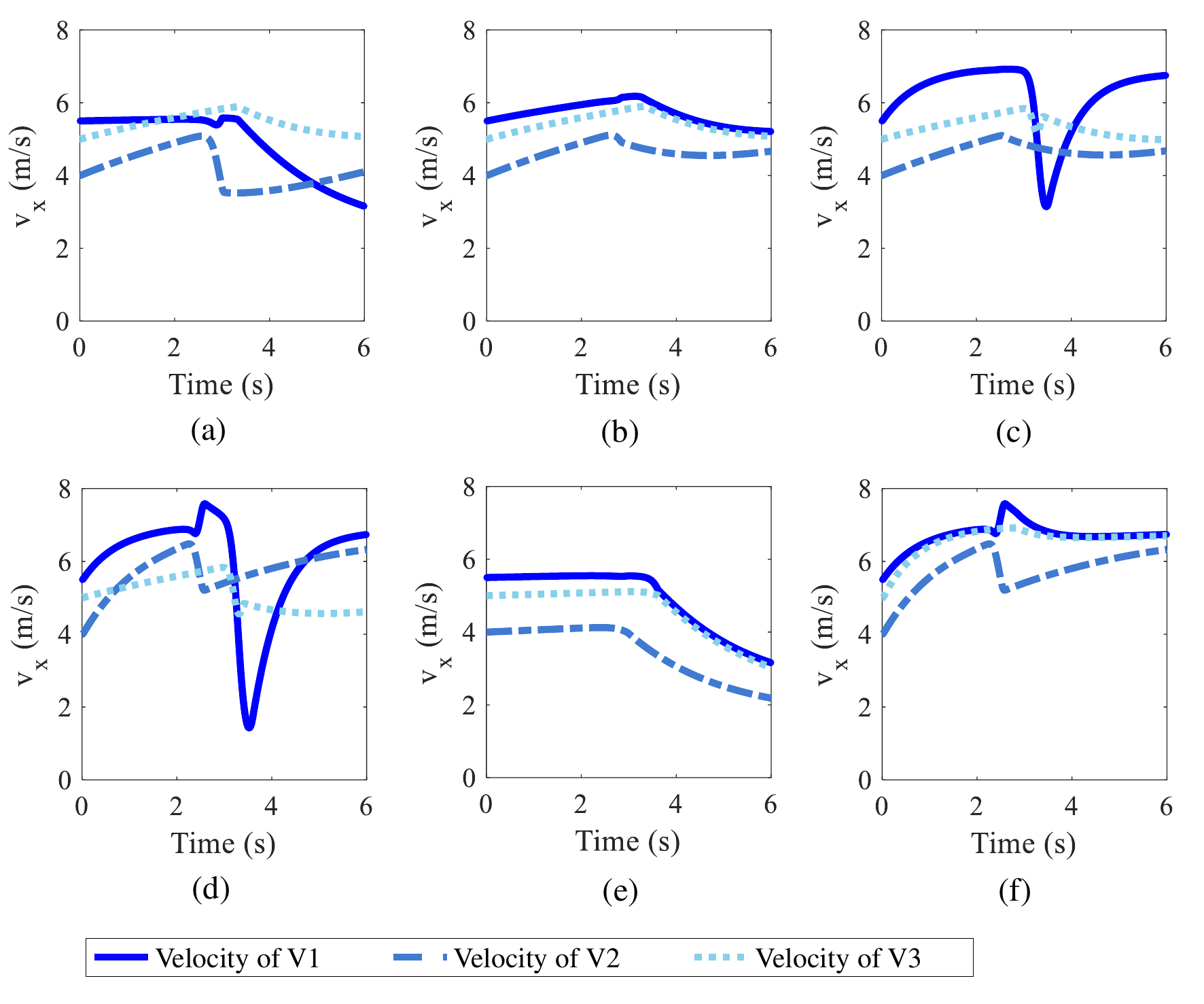}
\caption{Velocities of CAVs in Case 1: (a) Scenario A; (b) Scenario B; (c) Scenario C; (d) Scenario D; (e) Scenario E; (f) Scenario F.}\label{FIG_8}
\end{figure}

\begin{figure}[!t]\centering
	\includegraphics[width=6cm]{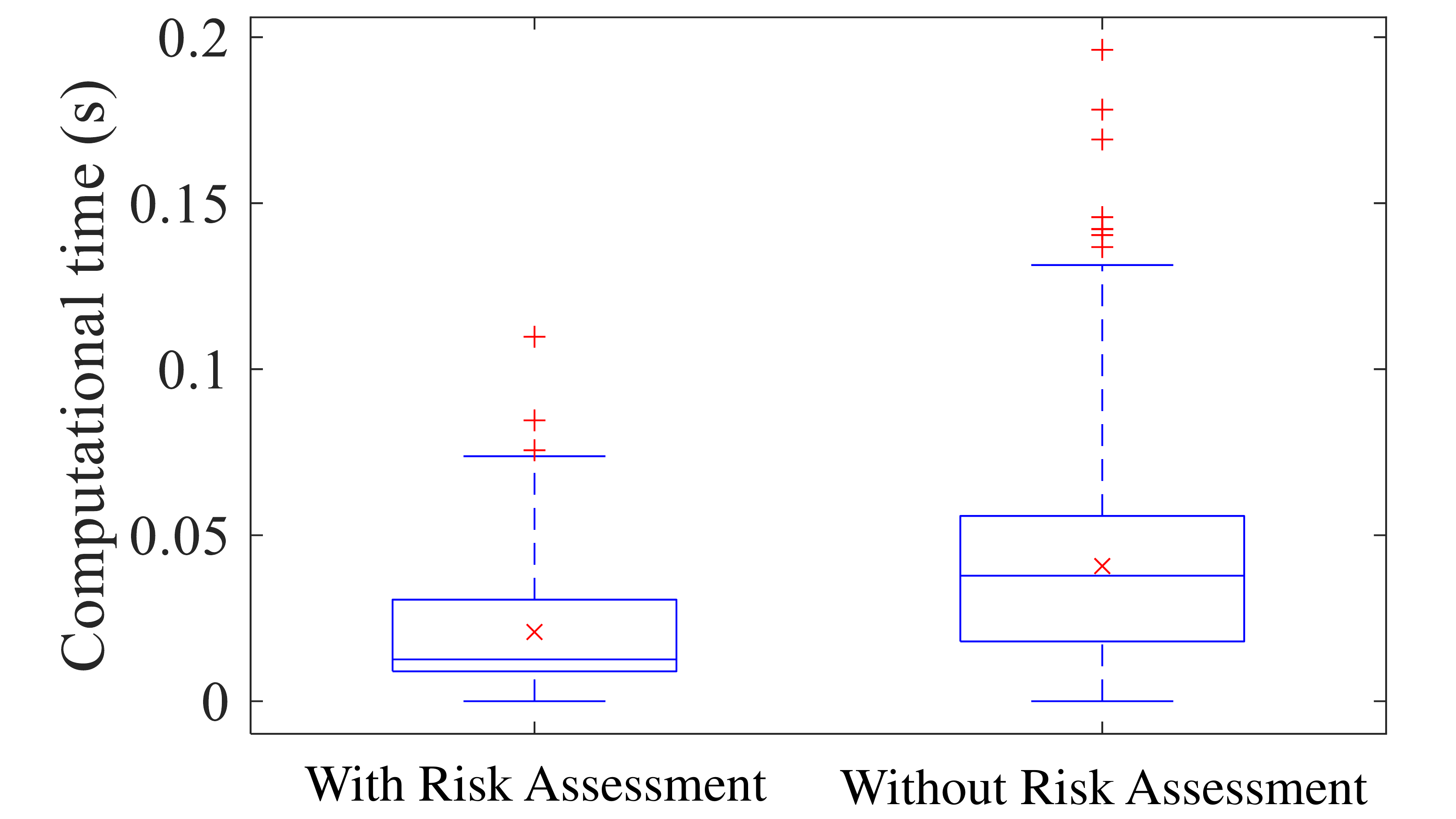}
\caption{Computational time at each time step in Case 1.}\label{Efficiency1}
\end{figure}

In Case 1, different driving aggressiveness coefficients are set for CAVs in the decision-making issue at an unsignalized intersection, which aims to verify the proposed algorithm that considers personalized decision making. In this case, V1 on the inside lane of the main road $M1$ wants to turn left and enter the inside lane of the main road $\hat{M}4$. At the conflict zone, it must interact with two conflict vehicles, V2 and V3, and then make the optimal decision. V2 on the inside lane of the main road $M2$ wants to turn left and enter the inside lane of the main road $\hat{M}1$. V3 on the inside lane of the main road $M3$ wants to move straight and enter the outside lane of the main road $\hat{M}1$.

To study the effect of personalized driving on decision making, six typical scenarios are designed according to different driving aggressiveness coefficients, which are shown in Table II. In this case, the initial position coordinates of V1, V2 and V3 are set as (-18, -2), (2, -15) and (20, 6), respectively. In addition, the initial velocities of V1, V2 and V3 are set as 5.5 m/s, 4 m/s and 5 m/s, respectively.

The test results of Case 1 are displayed in Figs. 5-7 and Tables III and IV. Figs. 5 and 6 show the motion paths and positions of the three CAVs at an unsignalized intersection, and Fig. 7 shows the motion velocities of all CAVs. The test results of the six different scenarios indicate that different driving aggressiveness coefficients lead to different decision-making results of CAVs. Generally, a larger $\kappa$ indicates a more aggressive driving behavior. In the decision-making process, the CAV with a larger $\kappa$ cares more about passing efficiency, which results in higher passing velocity. In contrast, a smaller $\kappa$ means more conservative driving behavior, which leads to a lower passing velocity. Scenarios E and F are two extreme conditions. In Scenario E, the three CAVs are all representative of conservative driving characteristics. However, all CAVs in Scenario F have high driving aggressiveness. Table III shows the maximum (Max) values and Root Mean Square (RMS) values regarding the velocity and acceleration of CAVs in Case 1.
The detailed comparative analysis presented in Table III reveals that regardless of whether a single CAV or the entire traffic system is considered, the velocities in Scenario F are much larger than those in Scenario E, which indicates that the passing efficiency of Scenario F is higher than that of Scenario E.
The test results of Scenarios C and D reveal that the high driving aggressiveness of CAVs will not deteriorate traffic efficiency, a benefit of the collaborative decision making of the fuzzy coalitional game. In contrast, it is impossible for human drivers.
Regarding driving safety, Table IV shows the test results of the relative distance and TTC. If two CAVs are highly aggressive (Scenarios D and F), it will result in a smaller relative distance and TTC, which is harmful to driving safety but within the safe boundary. In contrast, conservative driving characteristics contribute a larger safe distance and TTC to advance driving safety.
From the above analysis of the test results, it can be concluded that the fuzzy coalitional game approach can provide personalized decision making for CAVs and advance the performance of the entire traffic system.

Moreover, the algorithm efficiency is evaluated, which is shown in Fig. 8. Without the driving risk assessment algorithm, the mean value of the computational time at each time step is 0.0407s. However, with the application of the driving risk assessment algorithm, the mean value of the computational time at each time step is 0.0209s. The algorithm efficiency is remarkably improved.

\subsection{Case 2}

\begin{table*}[!t]
	\renewcommand{\arraystretch}{1.3}
	\caption{Velocities and accelerations of CAVs in Case 2}
\setlength{\tabcolsep}{5 mm}
	\centering
	\label{table_5}
	\resizebox{\textwidth}{18mm}{
		\begin{tabular}{c c c c c | c c c c | c c c c c c c c }
			\hline\hline \\[-4mm]
			\multirow{2}{*}{Test results} & \multicolumn{4}{c|}{Algorithm A} & \multicolumn{4}{c|}{Algorithm B} & \multicolumn{4}{c}{Algorithm C} \\
\cline{2-13} & \makecell [c] {V1} & \makecell [c] {V2} & \makecell [c] {V3} & \makecell [c] {V4} & \makecell [c] {V1} & \makecell [c] {V2} & \makecell [c] {V3} & \makecell [c] {V4} & \makecell [c] {V1} & \makecell [c] {V2} & \makecell [c] {V3} & \makecell [c] {V4} \\
\hline
			\multicolumn{1}{l}{Velocity Max / (m/s)} & 6.92 & 4.68 & 5.38 & 6.44 & 6.89 & 4.74 & 5.50 & 6.06 & 6.37 & 5.01 & 5.73 & 6.41\\
			\multicolumn{1}{l}{Velocity RMS / (m/s)} & 6.29 & 3.09 & 4.01 & 5.31 & 6.14 & 3.35 & 4.29 & 5.35 & 6.07 & 4.02 & 4.85 & 5.74\\
\multicolumn{1}{l}{Acceleration Max / $\mathrm{(m/s^2)}$} & 2.52 & 0.55 & 0.56 & 0.61 & 1.91 & 0.58 & 0.57 & 0.59 & 1.01 & 0.63 & 0.49 & 0.55\\
\multicolumn{1}{l}{Acceleration RMS / $\mathrm{(m/s^2)}$}  & 0.72 & 0.29 & 0.12 & 0.39 & 0.50 & 0.40 & 0.15 & 0.44 & 0.31 & 0.47 & 0.23 & 0.50\\
\multicolumn{1}{l}{Jerk Max / $\mathrm{(m/s^3)}$}  & 2.00 & 0.13 & 1.95 & 2.00 & 2.00 & 0.13 & 1.21 & 1.99 & 1.90 & 0.11 & 1.65 & 0.95\\
\multicolumn{1}{l}{Jerk RMS / $\mathrm{(m/s^3)}$}  & 0.28 & 0.03 & 0.11 & 0.23 & 0.26 & 0.03 & 0.08 & 0.14 & 0.25 & 0.04 & 0.07 & 0.11\\
\hline
\multicolumn{1}{l}{System Velocity RMS / (m/s)}  & \multicolumn{4}{c|}{4.83} & \multicolumn{4}{c|}{4.96} & \multicolumn{4}{c}{5.23} \\
			\hline\hline
		\end{tabular}
	}
\end{table*}

\begin{table*}[!t]
	\renewcommand{\arraystretch}{1.5}
	\caption{Relative Distance and TTC in Case 2}
\setlength{\tabcolsep}{5 mm}
	\centering
	\label{table_6}
	\resizebox{\textwidth}{10mm}{
		\begin{tabular}{c c c c | c c c | c c c }
			\hline\hline \\[-5mm]
			\multirow{2}{*}{Test results} & \multicolumn{3}{c|}{Algorithm A} & \multicolumn{3}{c|}{Algorithm B} & \multicolumn{3}{c}{Algorithm C} \\
\cline{2-10} & \makecell [c] {V1-V2} & \makecell [c] {V1-V3} & \makecell [c] {V1-V4} & \makecell [c] {V1-V2} & \makecell [c] {V1-V3} & \makecell [c] {V1-V4} & \makecell [c] {V1-V2} & \makecell [c] {V1-V3} & \makecell [c] {V1-V4} \\
\hline
			\multicolumn{1}{c}{Distance Min / (m)} & 5.41 & 4.63 & 4.21  & 5.25 & 4.45 & 4.20 & 4.60 & 3.72 & 8.16 \\
			\multicolumn{1}{c}{TTC Min / (s)} & 2.55 & 2.24 & 1.52  & 2.50 & 2.24 & 1.70 & 2.33 & 2.18 & 2.01 \\

			\hline\hline
		\end{tabular}
	}
\end{table*}

In Case 2, a comparative study among the proposed algorithm and other algorithms is carried out. Three game-theoretic decision-making algorithms are conducted, including a noncooperative game (Algorithm A), fuzzy coalitional game (Algorithm B), and cooperative grand coalitional game (Algorithm C).
In this case, V1 on the outside lane of the main road $M1$ wants to move straight and enter the outside lane of the main road $\hat{M}3$. At the conflict zone, it must interact with three conflict vehicles, V2, V3 and V4, and then make the optimal decision. V2 at the conflict zone wants to move straight and enter the outside lane of the main road $\hat{M}2$. V3 on the inside lane of the main road $M3$ wants to turn left and enter the inside lane of the main road $\hat{M}2$. V4 on the outside lane of the main road $M2$ wants to turn right and enter the outside lane of the main road $\hat{M}3$. The initial position coordinates of V1, V2, V3 and V4 are set as (-15, -6), (-6, 5), (10, 2) and (6, -20), respectively. Additionally, the initial velocities of V1, V2, V3 and V4 are set as 5.5 m/s, 4 m/s, 5 m/s and 4 m/s, respectively. The driving aggressiveness coefficients are set as 0.8, -0.1, -0.2 and 0.

It can be found that V1 must contend with four conflicts, including cross conflict with V1, cross conflict with V2, confluence conflict with V3, and following conflict with V3 after confluence. Both the quantity and type of driving conflict are greater than Case 1, which yields a more complex decision-making issue at an unsignalized intersection.

\begin{figure}[!t]\centering
	\includegraphics[width=8.5cm]{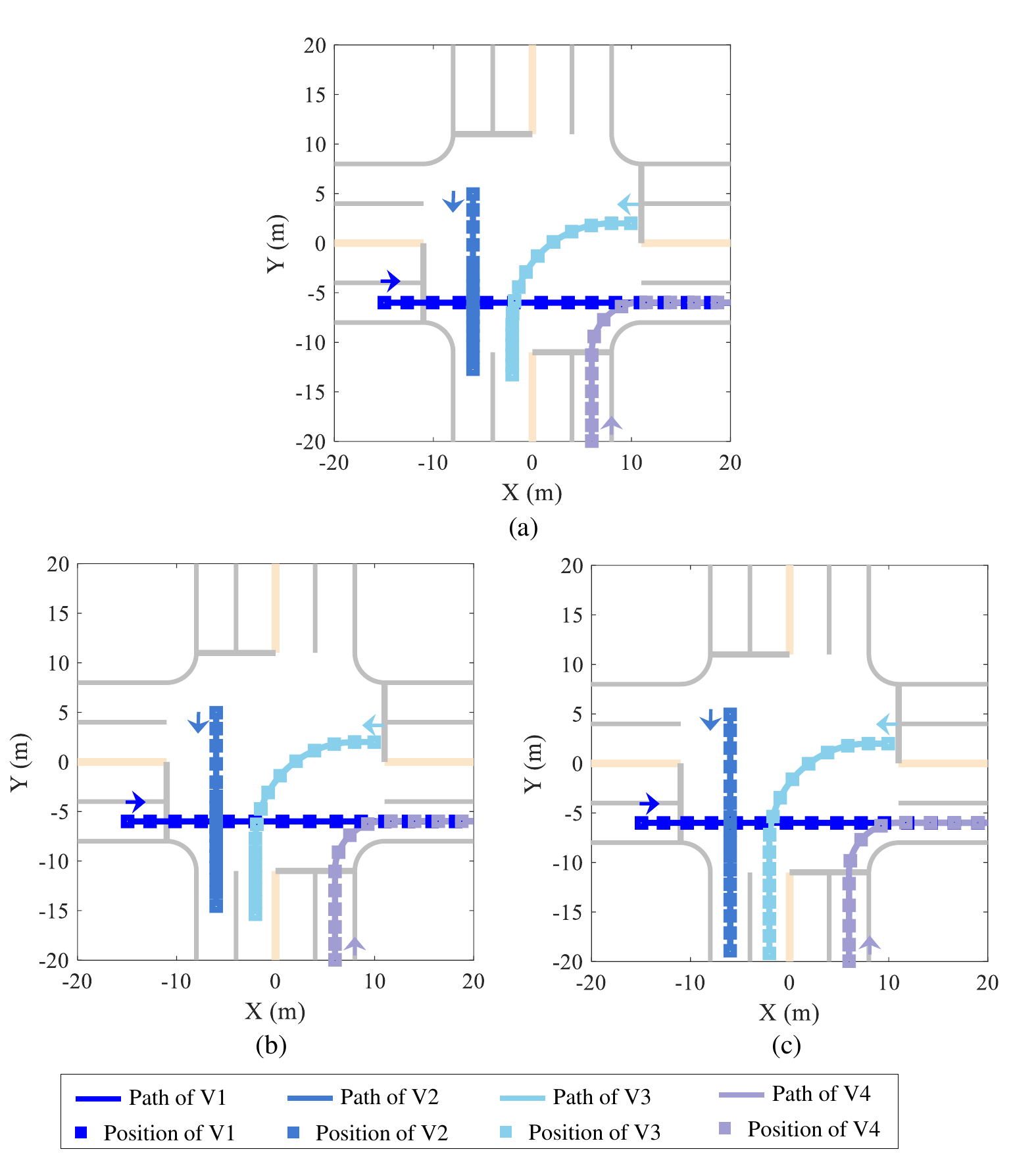}
\caption{Decision-making results of CAVs in Case 2: (a) Algorithm A; (b) Algorithm B; (c) Algorithm C.}\label{FIG_9}
\end{figure}

\begin{figure}[!t]\centering
	\includegraphics[width=8.5cm]{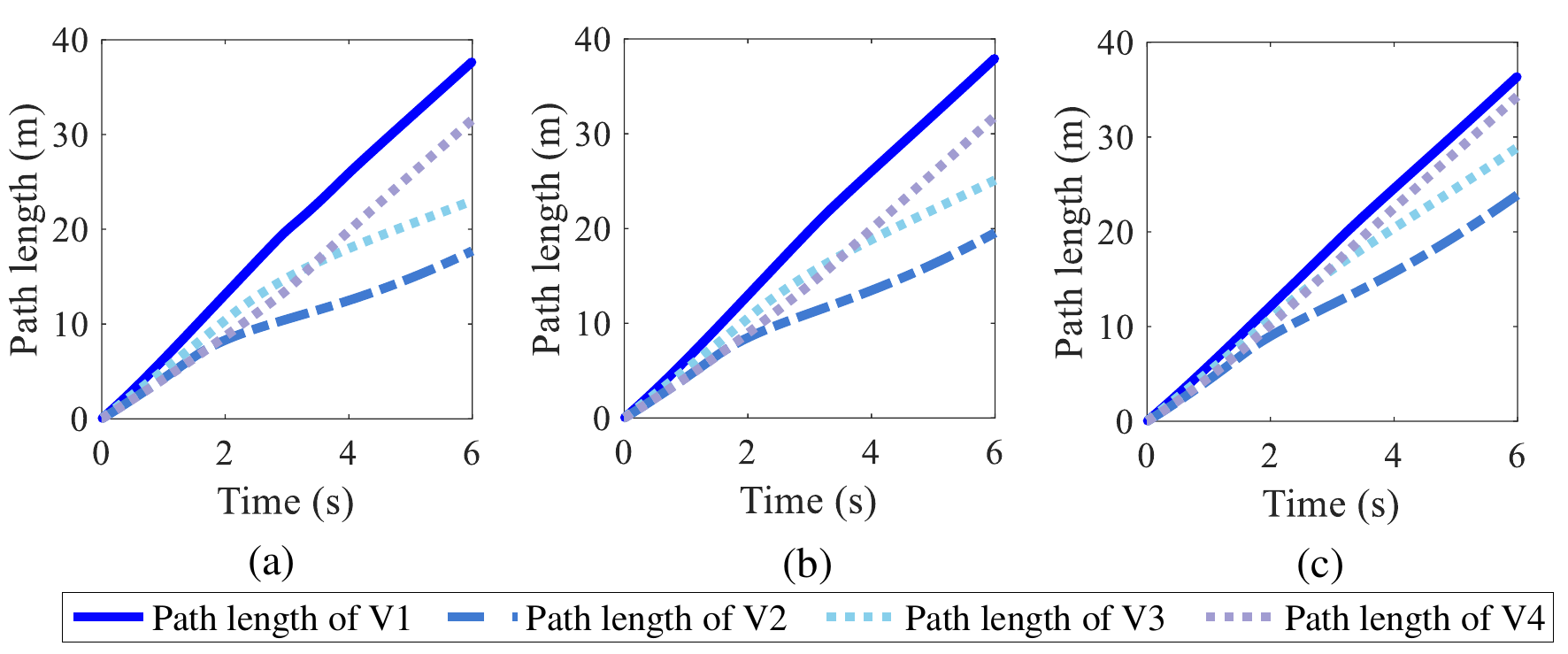}
\caption{Path length of CAVs in Case 2: (a) Algorithm A; (b) Algorithm B; (c) Algorithm C.}\label{FIG_10}
\end{figure}

\begin{figure}[!t]\centering
	\includegraphics[width=8.5cm]{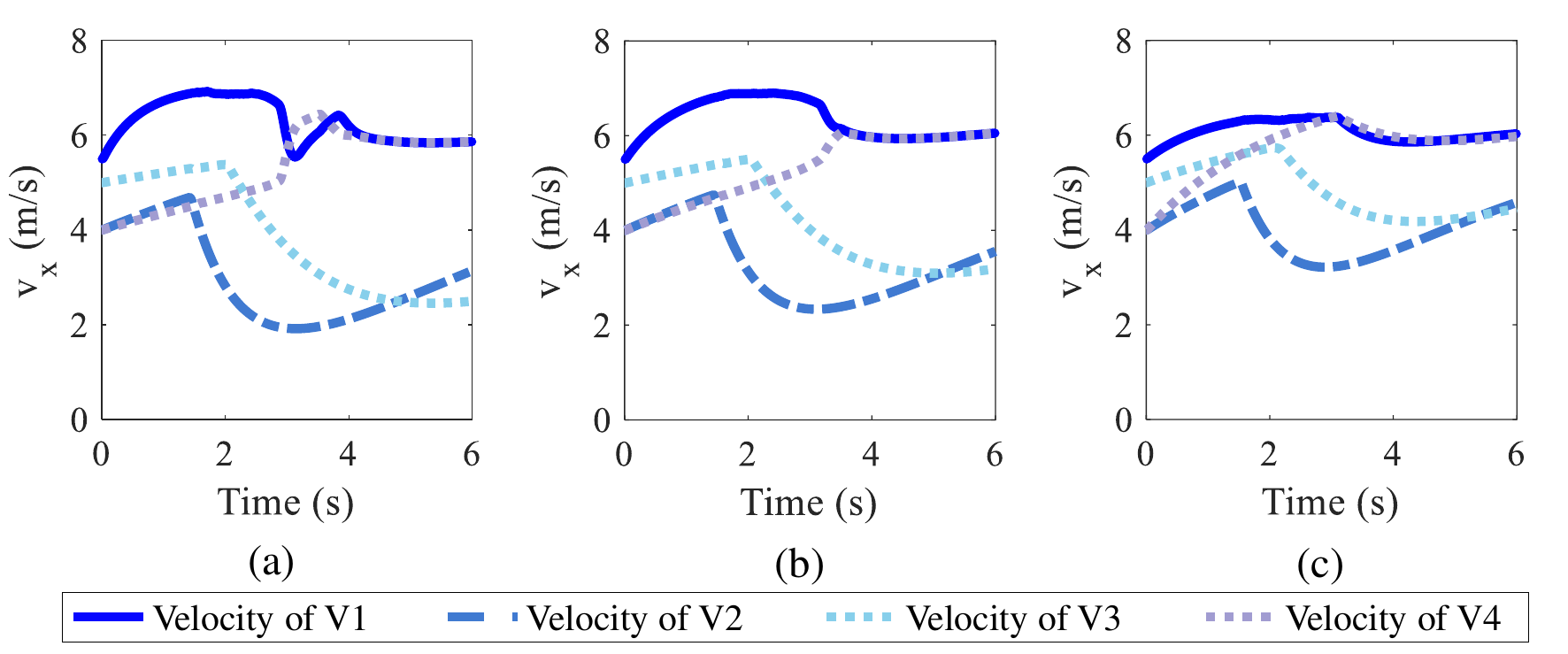}
\caption{Velocities of CAVs in Case 2: (a) Algorithm A; (b) Algorithm B; (c) Algorithm C.}\label{FIG_11}
\end{figure}

The test results of the three kinds of game-theoretic decision-making algorithms are presented in Figs. 9-11 and Tables V and VI. Figs. 9 and 10 show the motion paths and positions of CAVs at an unsignalized intersection, and Fig. 11 shows the motion velocities of CAVs. V1 has the largest velocity among the CAVs in all decision-making algorithms due to its highest driving aggressiveness. Additionally, we can see from Table V that V1 in Algorithm A has a larger velocity and acceleration than in the other two algorithms, which is decided by the principle of a noncooperative game. In the noncooperative game, each player only pursues its individual benefit rather than the social benefit of the traffic system. For V1, since it has higher driving aggressiveness, it cares more about passing efficiency, which results in its larger motion velocity, but the passing efficiency of the entire traffic system is worse. Algorithm C is based on the grand coalitional game, which aims to maximize the benefit of the entire traffic system. Therefore, it can be found that the system velocity of Algorithm C is largest among the three decision-making algorithms. However, the velocity of V1 is smallest, which indicates that the travel efficiency of the entire traffic system is increased, but personalized driving is sacrificed. Compared with the above two decision-making algorithms, Algorithm B, i.e., the fuzzy coalitional game approach, can both ensure personalized driving characteristics and improve traffic efficiency. Regarding driving safety, Table VI shows the test results of the minimum relative distance between CAVs and TTC. A similar conclusion can be drawn, reaffirming that the fuzzy coalitional game approach can make a good trade-off between individual benefits and social benefits for CAVs.

\subsection{Case 3}
\begin{table*}[!t]
	\renewcommand{\arraystretch}{1.3}
	\caption{Velocities and accelerations of CAVs in Case 3}
\setlength{\tabcolsep}{3 mm}
	\centering
	\label{table_3}
	\resizebox{\textwidth}{20mm}{
		\begin{tabular}{c c c c c c c c c | c c c c c c c c}
			\hline\hline \\[-4mm]
			\multirow{2}{*}{Test results} & \multicolumn{8}{c|}{Algorithm A} & \multicolumn{8}{c}{Algorithm B} \\
\cline{2-17} & \makecell [c] {V1} & \makecell [c] {V2} & \makecell [c] {V3} & \makecell [c] {V4} & \makecell [c] {V5} & \makecell [c] {V6} & \makecell [c] {V7} & \makecell [c] {V8} & \makecell [c] {V1} & \makecell [c] {V2} & \makecell [c] {V3} & \makecell [c] {V4} & \makecell [c] {V5} & \makecell [c] {V6} & \makecell [c] {V7} & \makecell [c] {V8}\\
\hline
			\multicolumn{1}{l}{Velocity Max / (m/s)} & 5.53 & 6.91 & 5.37 & 5.16 & 5.78 & 5.98 & 4.95 & 5.26 & 5.52 & 8.00 & 5.81 & 6.60 & 7.51 & 6.85 & 5.20 & 6.62\\
			\multicolumn{1}{l}{Velocity RMS / (m/s)} & 4.06 & 6.07 & 3.70 & 4.03 & 4.97 & 5.28 & 4.27 & 3.90 & 4.83 & 6.55 & 5.39 & 6.19 & 5.70 & 6.48 & 4.56 & 6.15 \\
\multicolumn{1}{l}{Acceleration Max / $\mathrm{(m/s^2)}$} & 8.00 & 2.82 & 1.01 & 0.54 & 5.13 & 1.38 & 6.69 & 1.10 & 6.04 & 8.00 & 0.86 & 1.46 & 7.61 & 3.16 & 1.17 & 2.10 \\
\multicolumn{1}{l}{Acceleration RMS / $\mathrm{(m/s^2)}$} & 1.54 & 0.61 & 0.23 & 0.41 & 0.78 & 0.36 & 0.92 & 0.58 & 0.85 & 1.78 & 0.32 & 0.45 & 1.33 & 0.82 & 0.50 & 0.77 \\
\multicolumn{1}{l}{Jerk Max / $\mathrm{(m/s^3)}$} & 2.00 & 2.00 & 2.00 & 0.30 & 2.00 & 2.00 & 2.00 & 0.29 & 2.00 & 2.00 & 1.36 & 0.18 & 2.00 & 2.00 & 2.00 & 0.87 \\
\multicolumn{1}{l}{Jerk RMS / $\mathrm{(m/s^3)}$} & 0.46 & 0.43 & 0.24 & 0.02 & 0.29 & 0.40 & 0.34 & 0.02 & 0.32 & 0.42 & 0.08 & 0.02 & 0.33 & 0.32 & 0.10 & 0.04 \\
\hline
\multicolumn{1}{l}{System Velocity RMS / (m/s)}  & \multicolumn{8}{c|}{4.60} & \multicolumn{8}{c}{5.77 $(\uparrow)$} \\
			\hline\hline
		\end{tabular}
	}
\end{table*}

\begin{figure}[!t]\centering
	\includegraphics[width=8.5cm]{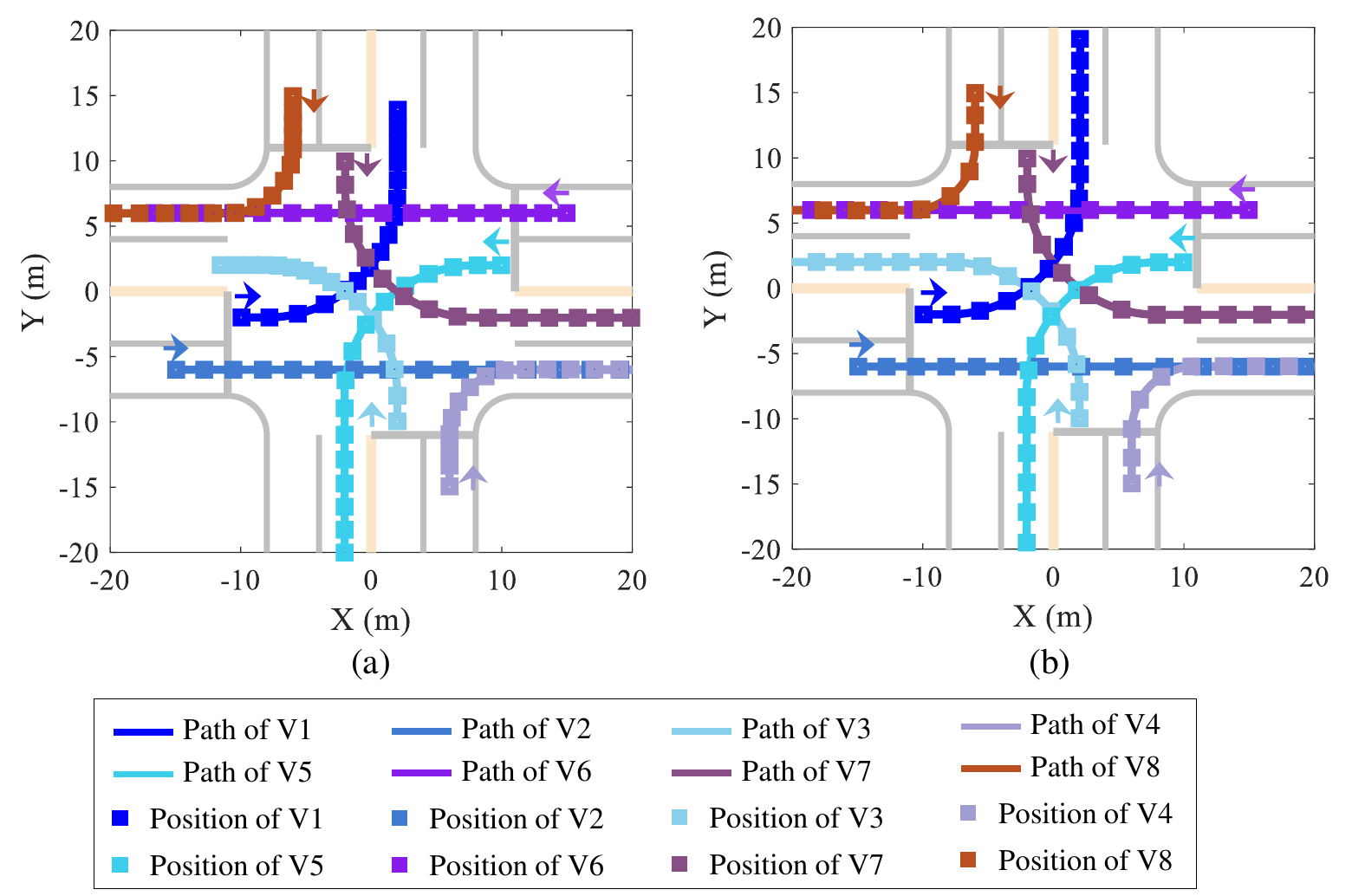}
\caption{Decision-making results of CAVs in Case 3: (a) Algorithm A; (b) Algorithm B.}\label{XY_Case3}
\end{figure}

\begin{figure}[!t]\centering
	\includegraphics[width=8.5cm]{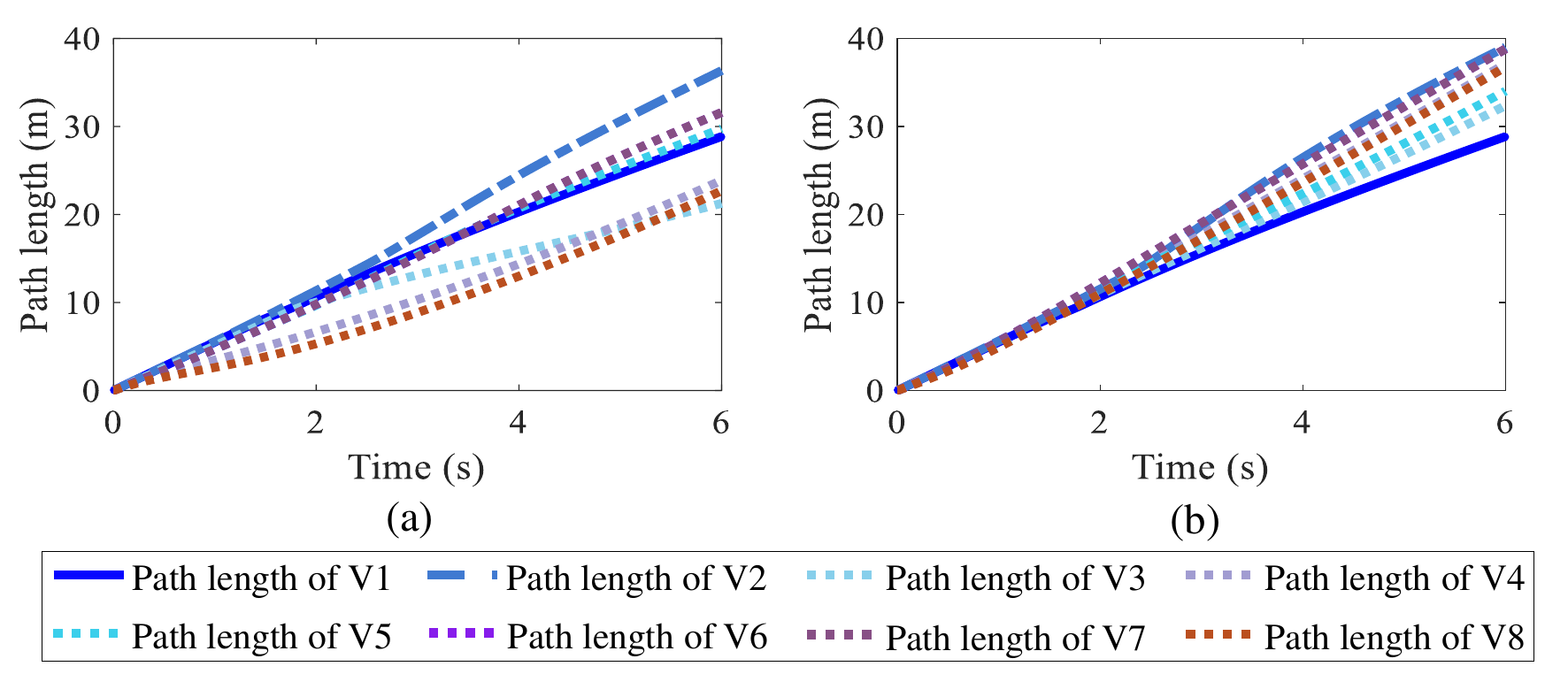}
\caption{Path length of CAVs in Case 3: (a) Algorithm A; (b) Algorithm B.}\label{Path_Case3}
\end{figure}

\begin{figure}[!t]\centering
	\includegraphics[width=8.5cm]{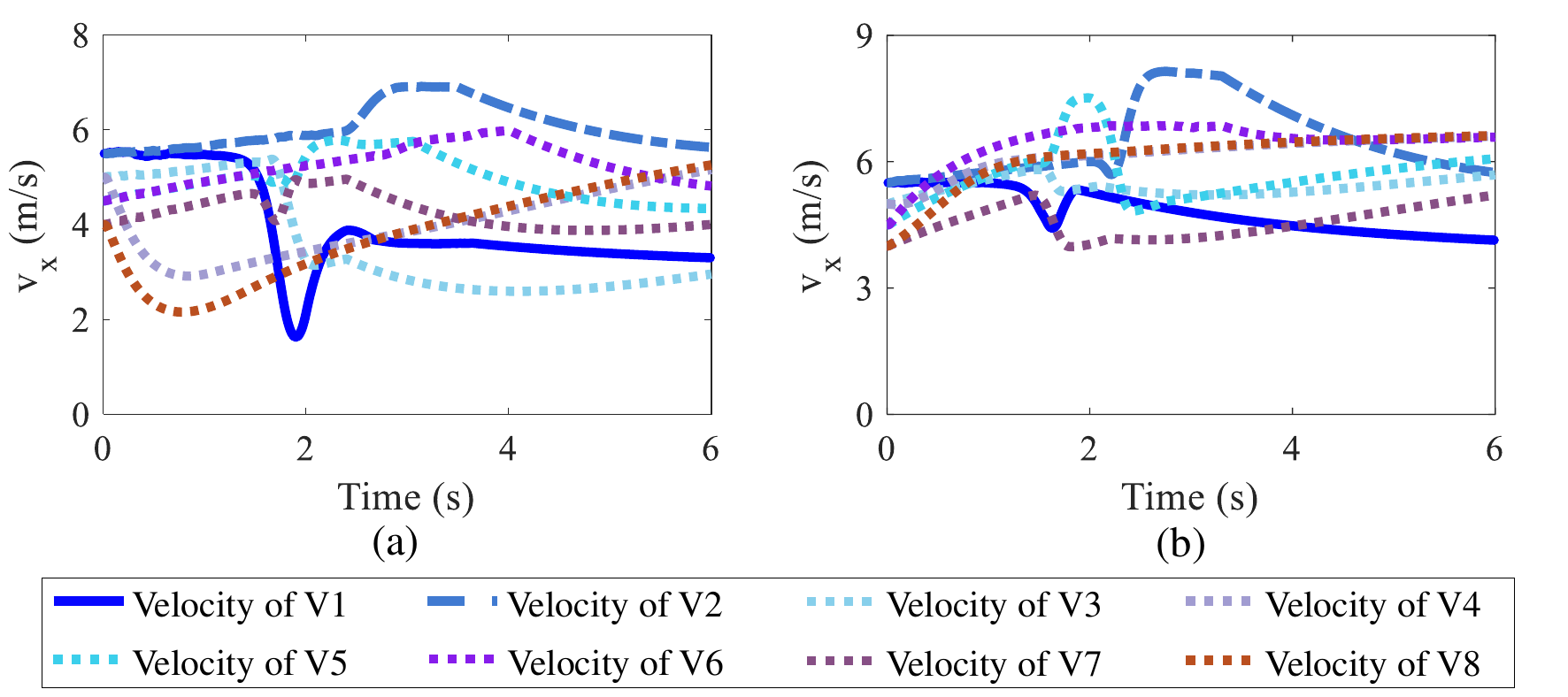}
\caption{Velocities of CAVs in Case 3: (a) Algorithm A; (b) Algorithm B.}\label{vx_Case3}
\end{figure}

\begin{table}[!t]
	\renewcommand{\arraystretch}{1.3}
	\caption{Relative Distance and TTC in Case 3}
\setlength{\tabcolsep}{0.8 mm}
	\centering
	\label{table_6}
	\resizebox{\columnwidth}{!}{
		\begin{tabular}{c c c | c c }
			\hline\hline \\[-5mm]
			\multirow{2}{*}{Test results} & \multicolumn{2}{c|}{Algorithm A} & \multicolumn{2}{c}{Algorithm B} \\
\cline{2-5} & \makecell [c] {Distance Min / (m)} & \makecell [c] {TTC Min / (s)} & \makecell [c] {Distance Min / (m)} & \makecell [c] {TTC Min / (s)} \\
\hline
			V1-V3 & 3.37 & 2.11 & 3.46 $(\uparrow)$ & 2.18 $(\uparrow)$\\
			V1-V6 & 4.43 & 1.50 & 4.64 $(\uparrow)$ & 1.52 $(\uparrow)$ \\
            V1-V7 & 3.60 & 1.50 & 4.41 $(\uparrow)$ & 1.51 $(\uparrow)$ \\
            V2-V3 & 7.49 & 3.70 & 7.84 $(\uparrow)$ & 3.96 $(\uparrow)$ \\
            V2-V4 & 3.45 & 2.25 & 13.31 $(\uparrow)$ & 3.82 $(\uparrow)$ \\
            V2-V5 & 4.40 & 2.22 & 3.38 $(\downarrow)$ & 1.77 $(\downarrow)$\\
            V3-V5 & 3.81 & 2.17 & 3.47 $(\downarrow)$ & 1.98 $(\downarrow)$\\
            V5-V7 & 3.56 & 2.12 & 3.04 $(\downarrow)$ & 1.91 $(\downarrow)$\\
            V6-V7 & 8.05 & 4.13 & 8.06 $(\uparrow)$ & 3.62 $(\downarrow)$\\
            V6-V8 & 6.88 & 1.61 & 13.81 $(\uparrow)$ & 3.37 $(\uparrow)$ \\
			\hline\hline
		\end{tabular}
	}
\end{table}

In Case 3, more vehicles are considered at the conflict zone, and a comparative study between the distributed conflict resolution mechanism \cite{liu2017distributed} (Algorithm A) and the fuzzy coalitional game approach (Algorithm B) is carried out. In this case, V1 on the inside lane of the main road $M1$ wants to turn left and enter the inside lane of the main road $\hat{M}4$. V2 on the outside lane of the main road $M1$ wants to move straight and enter the outside lane of the main road $\hat{M}3$. V3 on the inside lane of the main road $M2$ wants to turn left and enter the inside lane of the main road $\hat{M}1$. V4 on the outside lane of the main road $M2$ wants to turn right and enter the outside lane of the main road $\hat{M}3$. V5 on the inside lane of the main road $M3$ wants to turn left and enter the inside lane of the main road $\hat{M}2$. V6 on the outside lane of the main road $M3$ wants to move straight and enter the outside lane of the main road $\hat{M}1$. V7 on the inside lane of the main road $M4$ wants to turn left and enter the inside lane of the main road $\hat{M}3$. V8 on the outside lane of the main road $M4$ wants to turn right and enter the outside lane of the main road $\hat{M}1$.
The initial position coordinates of V1, V2, V3, V4, V5, V6, V7, and V8 are set as (-10, 2), (-15, -6), (2, -10), (6, -15),(10, 2), (15, 6), (-2, 10), and (-6, -15), respectively. Additionally, the initial velocities of V1, V2, V3, V4, V5, V6, V7, and V8 are set as 5.5 m/s, 5.5 m/s, 5 m/s, 5 m/s, 4.5 m/s, 4.5 m/s, 4 m/s, and 4 m/s, respectively. The driving aggressiveness coefficients are set as -0.2, 0.8, 0, 0.3, 0.2, 0.5, 0, 0.2, respectively. There exist ten driving conflicts for the eight vehicles.

The test results are illustrated in Figs. \ref{XY_Case3}-\ref{vx_Case3}. It can be found from Fig. \ref{XY_Case3} that there exist ten driving conflicts for the eight vehicles in this case. Two kinds of decision-making algorithms lead to different decision-making results, which are reflected by the motion velocities of CAVs in Fig. \ref{vx_Case3}. We can see that the motion velocities of most vehicles with Algorithm B are larger than that with Algorithm A. The analysis results in Table VII can support the conclusion that the travel efficiency for each individual CAV is improved with Algorithm B. Moreover, the system velocity of Algorithm B is also larger than that of Algorithm A, which means the passing efficiency of the entire traffic system is advanced as well. The performance analysis regarding safety is displayed in Table VIII. It can be found that the safe distance and TTC of most vehicles with Algorithm B are larger than that with Algorithm A, which means the driving safety can be guaranteed.

\begin{figure}[!t]\centering
	\includegraphics[width=6cm]{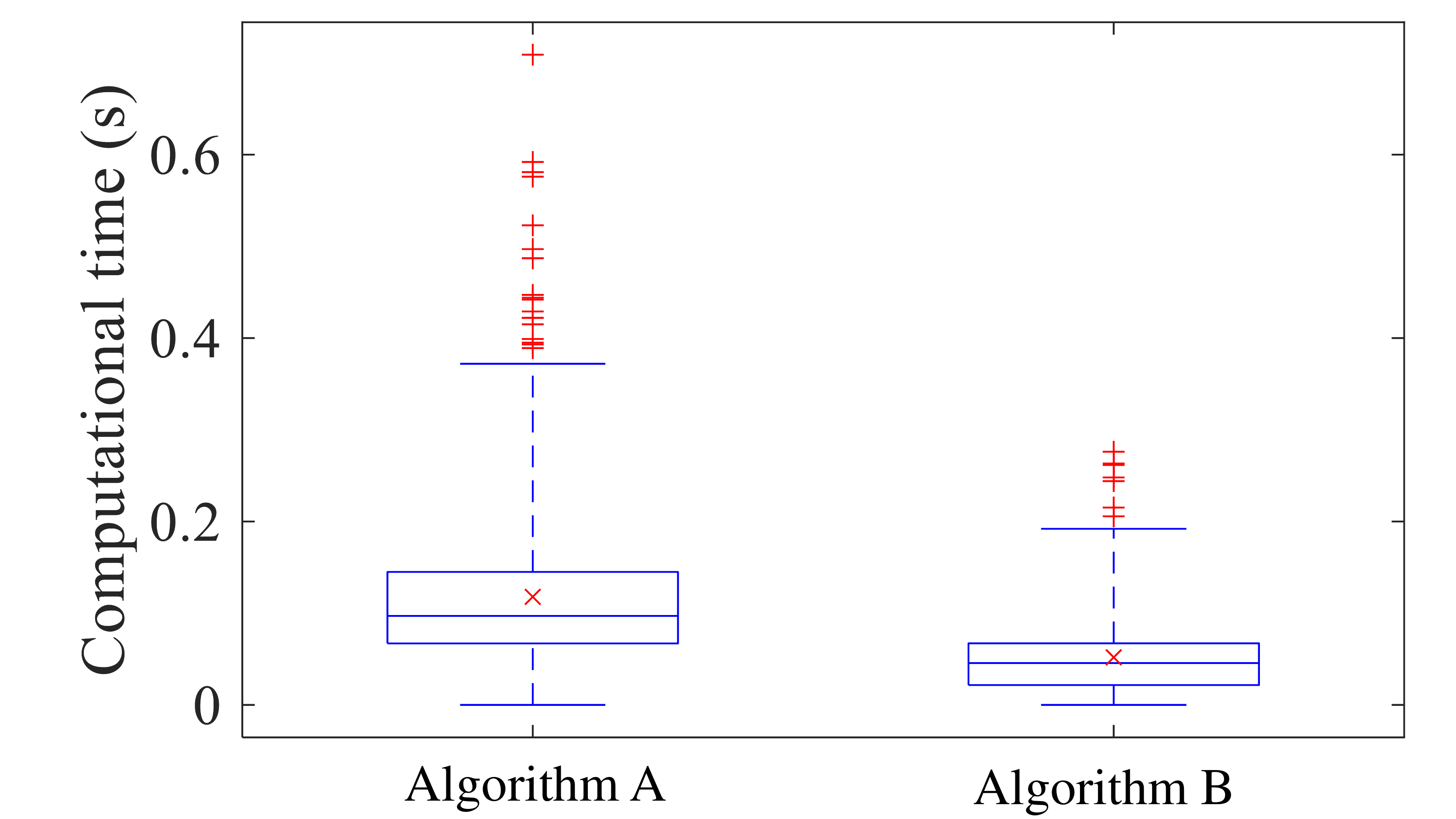}
\caption{Computational time at each time step in Case 3.}\label{Efficiency_Case3}
\end{figure}

Besides, the algorithm efficiency of the two decision-making algorithms are shown in Fig. \ref{Efficiency_Case3}. The mean values of the computational time at each time step for Algorithm A and Algorithm B are 0.1178s and 0.0519s, respectively. It indicates that the proposed fuzzy coalitional game approach has higher computational efficiency than the distributed conflict resolution mechanism. Compared with Case 1, although the number of CAVs increased from 3 to 8, the computational time has not shown a geometric increase due to the application of the driving risk assessment algorithm.

Based on the test results and analysis of the three cases, it can be concluded that the proposed fuzzy coalitional game approach can help CAVs make safe, reasonable and personalized decisions at unsignalized intersections. It integrates the advantages of both the noncooperative game approach and the grand coalitional game approach and can consider the individual and social benefits at the same time, which is beneficial to a single CAV and the entire traffic system.

\section{Conclusion}
In this paper, a game-theoretic decision-making framework is designed for CAVs to address the driving conflict encountered at unsignalized intersections. In this decision-making framework, both the individual benefit of a single CAV and the social benefit of the entire traffic system are considered due to the application of the fuzzy coalitional game approach. To enhance decision-making safety and improve algorithm efficiency, a Gaussian potential field approach is applied to the driving risk algorithm. In decision-making optimization, driving safety and passing efficiency are regarded as two key performances, and other performances, e.g., comfort, stability and energy, are set as constraint boundaries. Finally, three test cases are carried out to evaluate the effectiveness and feasibility of the proposed decision-making framework. Test results indicate that the decision-making framework is capable of making safe, efficient and reasonable decisions for CAVs. Additionally, it is in favor of the personalized driving of CAVs and improve the travel efficiency of the entire traffic system at unsignalized intersections.

Our future work will focus on the study of decision-making algorithms considering the interaction between CAVs and human drivers and further apply such algorithms to vehicle hardware platforms.

\ifCLASSOPTIONcaptionsoff
\newpage
\fi

\bibliography{Ref}

\end{document}